\documentclass[preprint,3p]{elsarticle}
\usepackage[T1]{fontenc}
\usepackage[utf8]{inputenc}
\usepackage{array}
\usepackage{float}
\usepackage{pdfcomment}
\usepackage{amsmath}
\usepackage{amssymb}
\usepackage{cancel}
\usepackage{graphicx}

\makeatletter

\providecommand{\tabularnewline}{\\}

\journal{Computer Physics Communications}

\usepackage{algpseudocode,algorithm,algorithmicx}
\usepackage{tikz}
\usetikzlibrary{shapes,arrows,calc,decorations.pathreplacing}
\usepackage{listings}
\usepackage{xcolor} 
\usepackage{booktabs}
\usepackage{rotating}
\lstdefinelanguage{YAML}{
  keywords={true,false,null,y,n},
  sensitive=false,
  comment=[l]{\#},
  morestring=[b]',
  morestring=[b]"
}
\ifdefined\showcaptionsetup
 \PassOptionsToPackage{caption=false}{subfig}
\fi
\usepackage{subfig}
\makeatother

\usepackage{babel}
\usepackage{listings}

\newcounter{bla}

\journal{Computer Physics Communications}

\hypersetup{
  hidelinks
}

\begin{document}
\begin{frontmatter}
\title{ChemGen: Code Generation for Multispecies Chemically Reacting Flow Simulations \author[NRL]{Ryan F. Johnson}
\author[NRL]{Eric~J.~Ching}
\author[SU]{Ethan~S.~Genter}
\author[SU]{Joshua~E.~Lipman}
\author[NRL]{Andrew~D.~Kercher}
\author[SDSU]{Jay~Arcities}
\author[SU]{Hai~Wang}
\address[NRL]{Laboratories for Computational Physics and Fluid Dynamics, U.S. Naval Research Laboratory, 4555 Overlook Ave SW, Washington, DC 20375}
\address[SU]{Department of Mechanical Engineering, Stanford University, Stanford, CA 94305}
\address[SDSU]{Department of Aerospace Engineering, San Diego State University, San Diego, CA 92182}}
\begin{abstract}
This paper introduces ChemGen, a software package that uses code generation
to integrate multispecies thermodynamics and chemical kinetics into
C\nolinebreak[4]\hspace{-.05em}\raisebox{.4ex}{\tiny\bf ++}-based
computational physics codes. ChemGen aims to make chemical kinetics
more accessible in existing simulation frameworks and help bridge
the gap between combustion modeling and computational physics. The
package employs the concept of decorators which enable flexible C\nolinebreak[4]\hspace{-.05em}\raisebox{.4ex}{\tiny\bf ++}
code generation to target established software ecosystems. ChemGen
generates code to evaluate thermodynamic properties, chemical source
terms, and their analytical derivatives for Jacobian calculations.
Also included are a variety of implicit time integration schemes,
linear solvers, and preconditioners. The various components of Chemgen
are verified by demonstrating agreement with Cantera and/or theoretical
convergence rates. Finally, we integrate ChemGen into OpenFOAM and
achieve a speedup over its native chemistry solver by approximately
four times. ChemGen is an ongoing project released under the NRL Open
License, a source-available license provided by the U.S. Naval Research
Laboratory.

\begin{small}
\noindent\\
{\em Program Title: ChemGen}                                          \\
{\em CPC Library link to program files:} N/A \\
{\em Developer's repository link: }\href{https://github.com/drryjoh/chemgen}{https://github.com/drryjoh/chemgen}
\\
{\em Licensing provisions: }CC0 1.0 \\
{\em Programming language: }Python and C\nolinebreak[4]\hspace{-.05em}\raisebox{.4ex}{\tiny\bf ++}       
\end{small}

\end{abstract}
\end{frontmatter}
\global\long\def\middlebar{\,\middle|\,}%
\global\long\def\average#1{\left\{  \!\!\left\{  #1\right\}  \!\!\right\}  }%
\global\long\def\expnumber#1#2{{#1}\mathrm{e}{#2}}%
 \newcommand*{\horzbar}{\rule[.5ex]{2.5ex}{0.5pt}} \def\Cpp{{C\nolinebreak[4]\hspace{-.05em}\raisebox{.4ex}{\tiny\bf ++ }}}

\global\long\def\revisionmathone#1{\textcolor{red}{#1}}%

\global\long\def\revisionmathtwo#1{\textcolor{blue}{#1}}%

\global\long\def\revisionmaththree#1{\textcolor{teal}{#1}}%

\makeatletter \def\ps@pprintTitle{  \let\@oddhead\@empty  \let\@evenhead\@empty  \def\@oddfoot{\centerline{\thepage}}  \let\@evenfoot\@oddfoot} \makeatother

\let\svthefootnote\thefootnote\let\thefootnote\relax\footnotetext{\\ \hspace*{65pt}DISTRIBUTION STATEMENT A. Approved for public release. Distribution is unlimited.}\addtocounter{footnote}{-1}\let\thefootnote\svthefootnote

\pdfcommentsetup{color=yellow}

\section{Introduction\label{sec:Introduction}}

Modeling of chemical reactions is critical to many areas of computational
physics. One of the most common examples is in modeling the  chemistry-turbulence coupling in Computational
Fluid Dynamics (CFD)~\citep{And95,Fer02,Hir07}, which is used to
simulate fluid behavior in a wide variety of reacting flows and combustion applications. The governing
equations, typically the Navier-Stokes equations, are discretized
in space using approaches such as finite volume~\citep{Pat80,Ver07,Mou16},
finite difference~\citep{LeV07}, and finite element schemes~\citep{Don03,Zie05}.
CFD relies on large volumes of data to describe the fluid; in particular,
the fluid state variables stored at all grid points in the computational
domain are known as the degrees of freedom (DoF).

Developing robust CFD software is often challenging due to computational challenges
such as efficient DoF communication in parallel, adaptive physical
modeling, and managing complex geometries. Additionally, there are
significant numerical issues, such as preserving stability~\citep{Kar94,Abg96,Bil03,Lv15,Lv17,Joh20,Chi22,Chi22_2}
which is coupled with the choice of formulation of the conservation
laws and time integration strategy. There is no single CFD software
that simultaneously addresses all the above difficulties for all fluid
dynamics applications, necessitating continued research into accelerating
simulations while achieving stable and accurate fluid modeling. As
a result, a broad family of CFD codes has emerged, ranging from experimental
in-house tools used for demonstrating new technologies to established
commercial software suites employed in production-level engineering.

A typical three-dimensional simulation of compressible, non-reacting
flow involves five conserved variables: energy, mass, and three momentum
components, such that $n_{y}=5$, where $n_{y}$ is the total state
size. In many cases, existing non-reacting CFD software built around
these five variables can be extended to chemically reacting flows
by expanding the state to include multiple chemical species and incorporating
$n_{r}$ chemical reactions. The state size then becomes $n_{y}=n_{s}+4$
, where $n_{s}$ represents the number of chemical species\footnote{The original five state variables are reduced by one since total density
can be determined from the species concentrations (or partial densities).}. Detailed chemical models may involve hundreds of species and thousands
of reactions, often with $n_{r}\sim5n_{s}$~\citep{Lu09}. The
number of species increases with the complexity of the fuel; for instance,
the hydrogen model in~\citep{Wes82,Chi22_2} consists of 10 chemical
species and 17 reactions, while the FFCM-2 model for simple hydrocarbons
(up to four carbons per chemical species) comprises 96 species and
1054 reactions~\citep{Zha23}.

In order to extend an existing single-species CFD software to basic
chemically reacting flows, three main features must be added to the
formulation:
\begin{enumerate}
\item \textbf{Multicomponent equation of state:} Most single-component formulations
assume a calorically perfect gas, where pressure is linear with respect
to internal energy. However, this assumption does not hold for multicomponent
reacting flow, requiring careful adjustments to ensure consistent
computations between among energy, temperature, and pressure.\label{enu:thermo}
\item \textbf{Stiff, nonlinear chemical source terms: }Other than external
forcing or simple bouyancy forces, the simplest CFD Navier-Stokes
models typically do not include a source term. In most models, momentum,
energy, and mass have no sources or sinks to add or remove them from
the domain; instead, the conserved variables are transported and diffused.
Furthermore, chemical source terms can be extremely stiff, introducing
additional numerical challenges.\label{enu:source}
\item \textbf{Multicomponent transport: }Transport quantities, such as viscosity
and thermal conductivity, would need to be adapted to multicomponent
mixture models. In addition, the diffusion of each species needs to
be considered with a transport method.
\end{enumerate}
There are several approaches to integrate these features. The most
common are:
\begin{enumerate}
\item Adapt existing legacy software, such as Chemkin~\citep{chemkin89},
for the target CFD code.\label{enu:chemkin}
\item Use the more widely adopted, open-source, and actively maintained
Cantera~\citep{cantera}, which can serve as a third-party library
for codes written in C, \Cpp, and Fortran, as well as in interpreted
languages such as Python and MATLAB.\label{enu:cantera}
\item Manually code chemical source term evaluations, thermodynamic fits,
transport properties, and other physical models.\label{enu:manual}
\item Use code generation tools to read in the chemical model and produce
adaptable source code in the desired format for the CFD software.\label{enu:software}
\end{enumerate}
Approaches~\ref{enu:chemkin} and~\ref{enu:cantera} are reasonable
choices but present certain challenges. With approach~\ref{enu:chemkin},
the existing legacy routines may not easily adapt to the specific
requirements of the targeted CFD software, limiting scalability and
complicating code maintenance. Regarding approach~\ref{enu:cantera},
although the package is excellent for property calculations , it can
only act as a third-party library, which may lead to limited control
over software functionality. Additionally, relying on a third-party
library(e.g. Cantera and Mutation++~\citep{Sco20}) may hinder compatibility
with targeted architectures that developers cannot directly access.
Using approach~\ref{enu:manual} can also introduce challenges; it
often requires extensive work by developers, as the necessary models
would need to be built from scratch (further described in Section~\ref{sec:governing_equations}).
This work belongs to approach~\ref{enu:software} and delivers a
tool to generate source code that is compatible with the existing
ecosystem of a target software package.

Code generation has a long history in chemical kinetics. Three public tools
focus on gas-phase combustion: PyJac~\citep{Nie17}, Pyrometheus~\citep{Cis25arxiv},
and KinetiX~\citep{Dan25}. PyJac generates analytical source-term Jacobians to
accelerate species evolution. Pyrometheus produces \Cpp{} headers or Python
modules for gas-phase kinetics with on-the-fly optimizations for scalable CFD.
KinetiX generates source terms, thermodynamic properties, and 
mixture-averaged transport models for CPU and GPU targets.  These tools broaden access to high-performance chemistry through code generation,
but, unfortunately, none of them provide a single framework that unifies source-term generation,
Jacobian construction within a differentiable code base, and stand-alone time
integration strategies, all of which are central to chemically reacting flow
simulations.

In this work, we introduce ChemGen, a software package that leverages
code generation to integrate multispecies thermodynamics, chemical
kinetics, and time integration strategies into \Cpp  based codes. ChemGen introduces decorators, discussed
in detail in Section~(\ref{sec:Code-Generation}), to enable flexible
\Cpp code generation tailored to existing software ecosystems. ChemGen
currently focuses on thermodynamic properties and chemical source
terms, as well as their analytical derivatives. The analytical derivatives
facilitate Jacobian assembly for implicit time integration strategies
appropriate for CFD. In addition, a number of time integrators, linear
solvers, and preconditioners is provided for convenience.

This paper is organized as follows. We begin with an overview of the
chemically reacting flow conservation equations used in CFD, highlighting
where chemistry fits within these formulations, including details
on multi-species thermodynamics and chemical reaction rates. Next,
we discuss time integration strategies, focusing on operator splitting
approaches that reduce the problem to a system of ordinary differential
equations for chemical state evolution, as well as fully coupled formulations.
We then introduce the code generation process, emphasizing how malleability
is achieved through the use of decorators. Finally, we present results
verifying the source term evaluation, Jacobian computation, and time
integration. We demonstrate ChemGen’s performance by replacing OpenFOAM’s
existing ODE-based chemical integrator in a detonation simulation
and achieve significantly faster time-to-solution due to ChemGen’s specialized,
generated source code. Overall, we show that ChemGen provides a path forward for improved malleability and computational efficiency compared to existing chemical kinetic solvers used in high-level CFD applications.

\section{Governing equations and thermodynamics \label{sec:governing_equations}}

In this work, we focus on a compressible CFD formulation without viscous
terms for simplicity, as we only seek to contextualize chemistry in
the framework of fluid dynamics. Here, we primarily highlight the
interaction between the fluid dynamic state, $y_{f}$, and the chemical
state, $y_{c}$.

The multicomponent chemically reacting compressible Euler equations
are given as
\begin{equation}
\frac{\partial y_{f}}{\partial t}+\nabla\cdot\mathcal{F}\left(y_{f}\right)-\mathcal{S}\left(y_{f}\right)=0\label{eq:conservation-law-strong-form}
\end{equation}
where $t$ is time, $y_{f}$ is the conservative state vector, $\mathcal{F}(y_{f})$
is the convective flux, and $\mathcal{S}(y_{f})$ is the CFD source
term that includes the chemical reactions. The fluid state is

\begin{equation}
y_{f}=\left(\rho v_{1},\ldots,\rho v_{d},\rho e_{t},C_{1},\ldots,C_{n_{s}}\right)^ {},\label{eq:reacting-navier-stokes-state}
\end{equation}
where $d$ is the spatial dimension, $n_{s}$ is the number of species
(which yields a state size $n_{y}=d+n_{s}+1$), $\rho$ is density,
$v=\left(v_{1},\ldots,v_{d}\right)$ is the velocity vector, $e_{t}$
is the mass-specific total energy, and $C=\left(C_{1},\ldots,C_{n_{s}}\right)$
is the vector of species molar concentrations. The density is computed
from the species concentrations as

\[
\rho=\sum_{i=1}^{n_{s}}\rho_{i}=\sum_{i=1}^{n_{s}}W_{i}C_{i},
\]

\noindent where $\rho_{i}$ is the partial density and $W_{i}$ is
the molecular mass of the $i^{th}$ species. The mass fraction of the
$i^{th}$ species is defined as 
\[
Y_{i}=\frac{W_{i}C_{i}}{\rho}=\frac{\rho_{i}}{\rho}.
\]

\noindent The $k$th spatial convective flux component is written
as
\begin{equation}
\mathcal{F}_{k}\left(y\right)=\left(\rho v_{k}v_{1}+p\delta_{k1},\ldots,\rho v_{k}v_{d}+p\delta_{kd},v_{k}\left(\rho e_{t}+p\right),v_{k}C_{1},\ldots,v_{k}C_{n_{s}}\right),\label{eq:reacting-navier-stokes-spatial-convective-flux-component}
\end{equation}
where $p$ is the pressure, and $\delta_{ij}$ is the Kronecker delta.
The mass-specific total energy is the sum of the internal and kinetic
energies, given by

\[
e_{t}=u+\frac{1}{2}\sum_{k=1}^{d}v_{k}v_{k},
\]
where the (mixture-averaged) mass-specific internal energy, $u$,
is:
\[
u=\sum_{i=1}^{n_{s}}Y_{i}u_{i}.
\]

The chemical state is

\begin{equation}
y_{c}=\left(T,C_{1},\ldots,C_{n_{s}}\right)\label{eq:reacting-navier-stokes-state-1-1}
\end{equation}
which is the complete derived state needed to calculate the components
of $\mathcal{S}\left(y_{f}\right)$. The species concentrations, $C_{i}$,
are all conserved quantities and can be extracted readily from the
fluid state along with the temperature, $T$. 

Beyond the chemical state values, several additional parameters are required. 
The first is the pressure, which may be obtained from the ideal gas law,
\begin{equation}
p = R T \sum_{i=1}^{n_{s}} C_{i},
\end{equation}
with $R$ is the universal gas constant. Nonideal gas treatment is possible for the equation of state and relevant thermodynamic properties, but this has not been attempted in the current version of ChemGen.
Another common parameter, used to reduce repeated logarithm evaluations in reaction optimizations, is $\ln T$. 
Finally, the mixture concentration,
\begin{equation}
M_{c} = \sum_{i=1}^{n_{s}} C_{i} = \frac{p}{R T},
\end{equation}
is required for evaluating pressure-dependent reaction rates. In what follows, we assume that the elements of $y_{c}$, together with $\ln T$, $p$, and $M_{c}$, 
are available throughout ChemGen. For brevity, functional dependencies are omitted.

\subsection{Thermodynamics}

In this section, we define the relevant thermodynamic quantities for
the $i^{th}$ species. The mass-specific
internal energy is given by~\citep{Gio99}
\begin{align}
u_{i}= & h_{i}-R_{i}T,\label{eq:internal_energy_identity}
\end{align}
where $h_{i}$ is the mass-specific enthalpy and $R_{i}=R/W_{i}$. The mass-specific enthalpy, $h_{i}$,
is given as
\begin{align}
h_{i}= & h_{\mathrm{ref},i}+\int_{T_{\mathrm{ref}}}^{T}c_{p,i}(\tau)d\tau,\label{eq:enthalpy_identity}
\end{align}
where $T_{\mathrm{ref}}=298\text{ K}$ is the reference temperature, $h_{\mathrm{ref},i}$ is the standard enthalpy of formation at $T_{\mathrm{ref}}$, and $c_{p,i}$ is the mass-specific heat capacity at constant pressure.
For differentiability and computational efficiency, ChemGen fits the
polynomial for $c_{p,i}$ to arbitrary order. As such, the value of
$c_{p,i}$ is computed from an $n_{p}$-order polynomial as
\begin{equation}
c_{p,i}=\sum_{k=0}^{n_{p}}a_{ik}T^{k},\label{eq:specific_heat_polynomial}
\end{equation}
which is based on, but not equivalent, to the NASA n-coefficient polynomial
parametrization~\citep{Mcb93,Mcb02}. Substituting Equation~(\ref{eq:specific_heat_polynomial})
into Equation~(\ref{eq:enthalpy_identity}), we evaluate the mass-specific
enthalpy as
\begin{align}
\begin{array}{ccc}
h_{i} & = & h_{\text{ref},i}+\sum_{k=0}^{n_{p}}\frac{a_{ik}}{k+1}T^{k+1},\\
 & = & \sum_{k=0}^{n_{p}+1}b_{ik}T^{k},
\end{array}\label{eq:enthalpy-polynomial}
\end{align}
with
\begin{equation}
b_{ik}=\begin{cases}
\frac{a_{i,k-1}}{k}, & k\geqq1\\
h_{\text{ref},i}, & k=0,
\end{cases}\label{eq:enthalpy_coeffs}
\end{equation}
where we have augmented the coefficients $a_{i}$ to include a coefficient
$h_{\text{ref},i}$, which corresponds to the zero degree polynomial,
i.e., $1$.

We specify $h_{\text{ref},i}$ as

\begin{equation}
h_{\text{ref},i}=\tilde{h}_{i}(T_{\mathrm{ref}}=298\text{ K})-\sum_{k=0}^{n_{p}}\frac{a_{ik}}{k+1}\left(T_{\mathrm{ref}}=298\text{ K}\right)^{k+1},\label{eq:enthalpy-adjustment}
\end{equation}
where $\tilde{h}_{i}\left(T_{\mathrm{ref}}=298\text{ K}\right)$ is
the value given by the NASA polynomials. This ensures that the species
enthalpies resulting from Equation~(\ref{eq:enthalpy-polynomial})
and those given by the NASA polynomials are equivalent when evaluated
at the reference temperature, $T_{\mathrm{ref}}=298\text{ K}$. Using
Equation~(\ref{eq:enthalpy-polynomial}), it is straight forward
to evaluate Equation~(\ref{eq:internal_energy_identity}) to obtain
the mass-specific internal energy. However, in practice, we define
the coefficients $c_{ik}$ by modifying the coefficients $b_{ik}$ to account for the term $R_{i}T$ and evaluate the mass-specific
internal energy as
\begin{align}
u_{i}= & \sum_{k=0}^{n_{p}+1}c_{ik}T^{k}.\label{eq:internal-energy-polynomial}
\end{align}
where

\begin{equation}
c_{ik}=\begin{cases}
\frac{a_{i,k-1}}{k}, & k>1\\
a_{i,0}-R_{i} & k=1\\
h_{\text{ref},i}, & k=0.
\end{cases}\label{eq:internal-energy-coeffs}
\end{equation}
Therefore, thermodynamic quantities can be evaluated by taking the
scalar inner product between the corresponding polynomial coefficients
and a monomial basis for temperature of the appropriate polynomial
degree, e.g., $n_{p}$ or $n_{p}+1$.

The mass-specific entropy of the mixture is defined
as
\begin{equation}
s=\sum_{i=1}^{n_{s}}Y_{i}s_{i},\label{eq:entropy-mass-definition}
\end{equation}
with $s_{i}$ given by
\begin{equation}
s_{i}=s_{i}^{o}-R\log\frac{p_{i}}{p_{\mathrm{ref}}},\quad s_{i}^{o}=s_{\mathrm{ref},i}^{o}+\int_{T_{\mathrm{ref}}}^{T}\frac{c_{p,i}(\tau)}{\tau}d\tau,\label{eq:species-entropy-definition}
\end{equation}
where $s_{\mathrm{ref},i}^{o}$ is the species entropy of formation
from the NASA n-coefficient polynomial parametrization evaluated at
$T=298$ K, and $p_{ref}$ is the reference pressure (usually $p_{\mathrm{ref}}=1\text{ atm}$),
$s_{i}^{o}$ denotes the mass-specific entropy at the reference pressure,
and $p_{i}=C_{i}R^ {}T$ is the partial pressure. The term, $R\log\frac{p_{i}}{p_{\mathrm{ref}}}$,
derives from the Gibbs equation, which accounts for the effect of pressure on entropy, and does not depend
on thermodynamic polynomial fits. Using Equation~(\ref{eq:specific_heat_polynomial}),
the mass-specific standard entropy, $s_{i}^{o}$, is computed as

\begin{equation}
s_{i}^{o}=s_{\mathrm{ref},i}^{o}+a_{i,0}\ln T+\sum_{k=1}^{n_{p}}\frac{a_{ik}}{k}T^{k}=a_{i,o}\ln T+\sum_{k=0}^{n_{p}}d_{ik}T^{k},\label{eq:entropy_polynomial}
\end{equation}
where

\begin{equation}
d_{ik}=\begin{cases}
\frac{a_{i,k}}{k}, & k\geq1\\
s_{\mathrm{ref},i}^{o} & k=0.
\end{cases}\label{eq:entropy-coeffs}
\end{equation}
Finally, the Gibbs free energy of the $i^{th}$ species is

\begin{equation}
g_{i}^{o}=h_{i}-Ts_{i}^{o}=b_{i0}-a_{i,0}T\ln T+\sum_{k=1}^{n_{p}}\left(b_{ik}-d_{i,k-1}\right)T^{k}+b_{i,n_{p}+1}T^{n_{p}+1},\label{eq:gibbs-reformatted}
\end{equation}
which can be represented as

\begin{equation}
g_{i}^{o}=-a_{i,0}T\ln T+\sum_{k=0}^{n_{p}+1}f_{ik}T^{k}.\label{eq:gibbs-polynomial}
\end{equation}
Note, the Gibbs free energy, like entropy, has a pressure dependency
that is calculated independent of the thermodynamic fits. Table~(\ref{tab:Summary-of-polynomial})
summarizes the thermodynamic relationships used by ChemGen.

\begin{table}[H]
\begin{tabular}{|>{\centering}p{1.35in}|>{\centering}p{1.95in}|>{\centering}p{2.6in}|}
\hline 
Thermodynamic Quantity of the $i^{th}$ Species & Fit & Coefficient Description\tabularnewline
\hline 
\hline 
Mass-specific heat at

constant pressure, $c_{p,i}$ & $c_{p,i}=\sum_{k=0}^{n_{p}}a_{ik}T^{k}$ & Re-fit from existing data~\citep{Mcb93,Mcb02}\tabularnewline
\hline 
Mass-specific enthalpy, $h_{i}$ & $h_{i}=\sum_{k=0}^{n_{p}+1}b_{ik}T^{k}$ & $b_{ik}=\begin{cases}
\frac{a_{i,k-1}}{k}, & k\geqq1\\
h_{\text{ref},i}, & k=0
\end{cases}$;

$h_{\text{ref},i}=h_{i}(T_{\mathrm{ref}})-\sum_{k=0}^{n_{p}}\frac{a_{ik}}{k+1}\left(T_{\mathrm{ref}}\right)^{k+1}$\tabularnewline
\hline 
Mass-specific internal energy, $u_{i}$ & $u_{i}=\sum_{k=0}^{n_{p}+1}c_{ik}T^{k}$ & $c_{ik}=\begin{cases}
\frac{a_{i,k-1}}{k}, & k>1\\
a_{i,0}-R & k=1\\
h_{\text{ref},i}, & k=0.
\end{cases}$\tabularnewline
\hline 
Mass-specific entropy at the reference pressure, $s_{i}^{o}$ & $s_{i}^{o}=a_{i,0}\ln T+\sum_{k=0}^{n_{p}}d_{ik}T^{k}$ & $d_{ik}=\begin{cases}
\frac{a_{i,k}}{k}, & k\geq1\\
s_{\mathrm{ref},i}^{o} & k=0.
\end{cases}$\tabularnewline
\hline 
Mass-specific Gibbs

free energy, $g_{i}^{o}$ & $g_{i}^{o}=-a_{i,0}T\ln T+\sum_{k=0}^{n_{p}+1}f_{ik}T^{k}$ & $f_{ik}=\begin{cases}
\left(\frac{1}{k}-\frac{1}{k-1}\right)a_{i,k-1} & n_{p}+1>k>1\\
a_{i,0}-s_{\mathrm{ref},i}^{o} & k=1\\
h_{\text{ref},i} & k=0.
\end{cases}$\tabularnewline
\hline 
\end{tabular}

\caption{Summary of polynomial coefficients in ChemGen for thermally perfect
species specific relationships.\label{tab:Summary-of-polynomial}}
\end{table}

The temperature of a given state, $y_f$, is defined such that the corresponding
volume specific internal energy,
\begin{equation}
\rho u=\frac{1}{2}\sum_{k=1}^{d}\frac{\left(\rho v_{k}\right)\left(\rho v_{k}\right)}{\sum_{i}^{n_{s}}W_{i}C_{i}}-\rho e_{t},\label{eq:internal-energy-density-state}
\end{equation}
which is consistent with the internal energy density computed via
\begin{equation}
\rho u(C,T)=\sum_{i}^{n_{s}}W_{i}C_{i}u_{i}\left(T\right),\label{eq:internal-energy-density-polynomial}
\end{equation}
where $u_{i}$ is the polynomial fit for mass specific internal energy
given by Equation~(\ref{eq:internal-energy-polynomial}). 

ChemGen solves the resulting nonlinear equation for temperature, namely,
\begin{equation}
\rho u-\rho u(C,T)=0,\label{eq:temperature-nonlinear}
\end{equation}
using Newton's method 
\begin{equation}
\frac{\partial\rho u(C,T)}{\partial T}\cdot\delta T=\rho u-\rho u(C,T).\label{eq:netwons-method-temperature}
\end{equation}
The Jacobian of $\rho u\left(C,T\right)$ with respect to temperature
is given by
\begin{align}
\begin{array}{ccc}
\frac{\partial\rho u(C,T)}{\partial T} & = & \sum_{i=1}^{n_{s}}W_{i}C_{i}\left(\sum_{k=1}^{n_{p}+1}c_{ik}\left(kT^{k-1}\right)\right),\\
 & = & \rho c_{v}\left(C,T\right),
\end{array}
\end{align}
where $c_{v}$ is the constant-volume specific heat, 

\begin{equation}
c_{v}=c_{p}-R.\label{eq:cv_equation}
\end{equation}
Therefore, temperature at the $k\text{th}$ Newton iteration is given
by

\begin{equation}
T_{k}=T_{k-1}-\frac{\left(\rho u-\rho u(C,T_{k-1})\right)}{\rho c_{v}\left(C,T_{k-1}\right)}\;\;\textrm{for }k=1,\ldots,n,\label{eq:netwon_iteration_temperature}
\end{equation}
where $n$ is the number of iterations and $T_{0}$ is the initial
guess for temperature. In ChemGen, we
set $n=5$, which, in our experience, is sufficient to be within $1\times10^{-8}$
relative error for a variety of models at different conditions.

\subsection{Chemical reaction rates and their ChemGen representations\label{subsec:chemical-reaction-rates}}

The source term in Equation~(\ref{eq:conservation-law-strong-form})
is a function of the fluid state, written as~\citep{Kee96}

\begin{equation}
\mathcal{S}\left(y_{f}\right)=\left(0,\ldots,0,0,\omega_{1},\ldots,\omega_{n_{s}}\right),\label{eq:reacting-navier-stokes-source-term}
\end{equation}
where $\omega_{i}$ is the production rate of the $i^{th}$ specie, which
satisfies mass conservation:
\begin{equation}
\sum_{i=1}^{n_{s}}W_{i}\omega_{i}=0.\label{eq:chemical-reaction-mass-conservation}
\end{equation}
For an arbitrary reaction model written as

\begin{equation}
\sum_{i=1}^{n_s}\nu_{ij}^{f}A_{i}\rightleftharpoons \sum_{i=1}^{n_s}\nu_{ij}^{r}A_{i}\ (j = 1,...n_r),
\end{equation}
where $A_i$ refers to the $i^{th}$ species $\nu_{ij}$ refers the stoichiometric coefficient of species $i$ in reaction $j$ with superscript $f$ and , the production rate is computed as
\[
\omega_{i}=\sum_{j=1}^{n_{r}}\nu_{ij}q_{j}.
\]
In the above equation, $n_{r}$ is the number of reactions, $\nu_{ij}=\nu_{ij}^{r}-\nu_{ij}^{f}$
is the difference between the reverse ($\nu_{ij}^{r}$) and the forward
stoichiometric coefficients ($\nu_{ij}^{f}$), and $q_{j}$ is the
rate of progress of the $j^{th}$ reaction, computed as
\begin{equation}
q_{j}=k_{j}^{f}\prod_{i=1}^{n_{s}}C_{i}^{\nu_{ij}^{f}}-k_{j}^{r}\prod_{i=1}^{n_{s}}C_{i}^{\nu_{ij}^{r}},\label{eq:chemical-reaction-rate-of-progress}
\end{equation}
where $k_{j}^{f}$ and $k_{j}^{r}$ are the forward and reverse rate
constants, respectively, of the $j^{th}$ reaction. The forward and reverse
rate constants are related via the equilibrium constant

\begin{equation}
K_{j}=\frac{k_{j}^{f}}{k_{j}^{r}},\quad k_{j}^{r}=\frac{k_{j}^{f}}{K_{j}},\label{eq:reverse_rate}
\end{equation}
where
\begin{equation}
K_{j}^ {}=\exp\left(-\frac{\Delta G_{j}'}{RT}\right)\left(\frac{p_{\mathrm{ref}}}{R^ {}T}\right)^{\sum_{i}\nu_{ij}},\label{eq:equilibrium-constant-1}
\end{equation}
with $\Delta G_{j}'$ denoting the change in reference-state Gibbs
free energy for the $j^{th}$ reaction, given as
\[
\Delta G_{j}'=\sum_{i=1}^{n_{s}}\nu_{ij}g_{i}^{o}.
\]
For the term $\left(\frac{p_{ref}}{RT}\right)^{\sum_{i}\nu_{ij}}$in~(\ref{eq:equilibrium-constant-1}),
ChemGen inspects $\sum_{i}\nu_{ij}$ and finds the specific power.
If the power is a positive integer $\left(\frac{p_{\mathrm{ref}}}{R^ {}T}\right)$
is expressed as a series of multiplications, e.g. $\sum_{i}\nu_{ij}=3$
gives $\left(\frac{p_{\mathrm{ref}}}{RT}\right)\cdot\left(\frac{p_{\mathrm{ref}}}{RT}\right)\cdot\left(\frac{p_{\mathrm{ref}}}{RT}\right)$
with $1/RT$ precomputed.Similarly, if $\sum_{i}\nu_{ij}$ is a negative
integer, ChemGen precomputes $1/p_{\mathrm{ref}}$ and then calculates
the corresponding power of $RT$, e.g., $\sum_{i}\nu_{ij}=-3$ gives
$\left(1/p_{\mathrm{ref}}\right)^{3}\left(RT\right)^{3}$ . Without
code generation, these types of expressions require expensive exponentiation
functions and possibly if-statements .

There exist various models for approximating the forward rate expressions,
$k_{j}^{f}$, in Equation~(\ref{eq:chemical-reaction-rate-of-progress}).
The reaction models currently supported by ChemGen are discussed below.

\subsubsection{Arrhenius rate expression}

The Arrhenius form is the most common model for approximating reaction
rates. The forward rate constants are computed as
\[
k_{j}^{f}=k_{A,j}^{f}=A_{j}T^{b_{j}}\exp\left(-\frac{E_{j}}{R^ {}T}\right),
\]
 where $A_{j}$ is the prefactor, $b_{j}$ is the temperature exponent,
and $E_{j}$ is an quasi-activation energy~\citep{Gio99,Kee96}. In ChemGen,
an optimization is made to again avoid the power function. Generally,
$b_{j}$ is not an integer, however, since the exponential function
is unavoidable, we can utilize the precomputed $\ln T$ and insert
an additional term in the exponent

\[
k_{A,j}^{f}=A_{j}\exp\left(b_{j}\ln T-\frac{E_{j}}{R^ {}T}\right).
\]

\subsubsection{Pressure-dependent reactions}

Evaluation of rates of pressure-dependent reactions requires the use of the mixture concentration,
which represents the collective effect of all species acting as a
third body in collisional activation or deactivation. Dissociation and recombination reactions are often of
this type. The rate of progress is scaled by a prefactor as~\citep{Kee96}
\begin{equation}
q_{j}=\left(\sum_{i=1}^{n_{s}}\alpha_{ij}C_{i}\right)\left(k_{j}^{f}\prod_{i=1}^{n_{s}}C_{i}^{\nu_{ij}^{f}}-k_{j}^{r}\prod_{i=1}^{n_{s}}C_{i}^{\nu_{ij}^{r}}\right),\label{eq:third_body}
\end{equation}
where $\alpha_{ij}$ are the third-body efficiencies. In ChemGen,
the rate of progress is treated to always resemble Equation ~(\ref{eq:chemical-reaction-rate-of-progress});
therefore, the prefactor for third-body reactions is incorporated
into the Arrenhius expression as

\begin{equation}
k_{j}^{f}=k_{TB,j}^{f}=\left(\sum_{i=1}^{n_{s}}\alpha_{ij}C_{i}\right)k_{A,j}^{f}=\left(M_{c}+\sum_{i=1}^{n_{s}}\left(\alpha_{ij}-1\right)C_{i}\right)k_{A,j}^{f}\label{eq:TB_optimization}
\end{equation}
which is still attributed to the second term in Rquation ~(\ref{eq:third_body})
via the relationship of $k_{j}^{f}$ to $k_{j}^{r}$ from Equation~(\ref{eq:reverse_rate}).
An additional optimization is made during the code generation process
for the $\alpha_{ij}$ terms. The established convention is that the
efficiencies always default $\alpha_{ij}=1$. Therefore, for any species
without an explicitly prescribed third-body or Chaperone efficiency efficiency, the $\left(\alpha_{ij}-1\right)C_{i}$
term is removed during code generation. This avoids possibly repetitive
work to determine the Chaperon efficiency from a sum of products of all species
to a sum of products of all species with $\alpha_{ij}\ne1$ by utilizing
the precomputed $M_{c}$.

\subsubsection{Pressure fall-off effect\label{subsec:fall-off-reactions}}

Unimolecular/recombination reactions in the pressure fall-off region require special consideration. Given Arrhenius-type low-pressure and
high-pressure limits for the rate functions ($k_{A,0,j}$ and $k_{A,\infty,j}$,
respectively with their own $A_{j}$, $b_{j}$, and $E_{j}$ constants),
$k_{j}^{f}$ is computed as
\begin{equation}
k_{j}^{f}=k_{FO,j}^{f}=k_{A,\infty,j}\left(\frac{P_{r}}{1+P_{r}}\right)F,\label{eq:falloff-reaction}
\end{equation}
where $P_{r}$ is the reduced pressure, defined as
\[
P_{r}=\frac{k_{A,0,j}}{k_{A,\infty,j}}\left(M_{c}+\sum_{i=1}^{n_{s}}\left(\alpha_{ij}-1\right)C_{i}\right).
\]
The same optimization for efficiencies can be made as in the third-body
prefactor. Various falloff reactions exist where the only difference
is in the function $F$, known as the broadening term, which we
outline below, the simplest being $F=1$ from Lindemann~\citep{Lin22}.

\subsubsection*{Troe broadening}

For the Troe fall-off reaction the broadening term, $F$, in Equation~(\ref{eq:falloff-reaction})
is computed using the following relations:

\begin{eqnarray}
\log_{10}F & = & \frac{\log_{10}F_{cent}}{1+f_{1}^{2}}\label{eq:f_and_fcent}\\
F_{cent} & = & (1-\alpha_{j})\exp(-T/T_{j,3})+\alpha\exp(-T/T_{j,1})+\exp(-T_{j,2}/T)\nonumber \\
f_{1} & = & (\log_{10}P_{r}+c)/(n-0.14(\log_{10}P_{r}+c))\nonumber \\
c & = & -0.4-0.67\;\log_{10}F_{cent}\nonumber \\
n & = & 0.75-1.27\;\log_{10}F_{cent}\nonumber 
\end{eqnarray}
The origin of the empirical parameters is given in~\citep{Tro77,Tro79,Gil83,Tro87}.

\subsubsection*{SRI broadening}

In the SRI~\citep{Ste89} broadening approach, $F$ is represented
by

\[
F=d\bigl[a\exp(-b/T)+\exp(-T/c)\bigr]^{1/(1+\log_{10}^{2}P_{r})}T^{e},
\]
where coefficients $a$ through $e$ are specified for each reaction.

\subsubsection{Other forms of pressure-dependent reactions}

Some reaction rates have several pressure-dependent expressions that cannot be accurately represented
by single Arrhenius expressions. Known as the P-log reaction rate, logarithmically interpolating between Arrhenius rate expressions at various pressures~\citep{Gou11_b} is required.  Specifically, 
given two Arrhenius rates at two specific pressures,

\begin{eqnarray}
p_{i}:k_{i} & = & A_{i}T^{b_{i}}\exp\left(-E_{i}/RT\right)\text{ and }\label{eq:plog_p1_p2}\\
p_{i+1}:k_{i+1} & = & A_{i+1}T^{b_{i+1}}\exp\left(-E_{i+1}/RT\right),\nonumber 
\end{eqnarray}
the rate at an intermediate pressure is computed as

\begin{equation}
\ln k_{plog}=\ln k_{i}+\bigl(\ln k_{i+1}-\ln k_{i}\bigr)\frac{\ln p-\ln p_{i}}{\ln p_{i+1}-\ln p_{i}}.\label{eq:plog_rate}
\end{equation}
An additional complexity is that multiple rate expressions may be
given at the same pressure because of non-Arrhenius behavior, in which case the rate used in the interpolation
formula is the sum of all the rates given at that pressure. For pressures
outside the given range, the rate expression at the nearest bounded
pressure is used. In ChemGen, the $\ln p_{i}$ terms are calculated
during code generation. 

To further elaborate on this rate, consider, for example, the 81st
reaction in FFCM2~\citep{Zha23},

\begin{lstlisting}[language=YAML, numbers=left, numberstyle=\tiny, stepnumber=1,
  frame=single, breaklines=true, basicstyle=\ttfamily\small]
- equation: CH2(S) + H2O <=> CH2O + H2  # Reaction 81
  type: pressure-dependent-Arrhenius
  rate-constants:
  - {P: 0.1 atm, A: 3.36e+22, b: -3.33, Ea: 3950.0}
  - {P: 1.0 atm, A: 4.8e+23, b: -3.63, Ea: 5220.0}
  - {P: 3.0 atm, A: 6.85e+23, b: -3.66, Ea: 5820.0}
  - {P: 10.0 atm, A: 1.53e+24, b: -3.73, Ea: 6820.0}
  - {P: 30.0 atm, A: 8.75e+23, b: -3.62, Ea: 7655.0}
  - {P: 100.0 atm, A: 1.27e+22, b: -3.06, Ea: 7950.0},
\end{lstlisting}

which requires seven boolean conditionals to be checked, $p\leq0.1$,
$0.1<p\leq1.0$, $1.0<p\leq3.0$, $3.0<p\leq10.0$, $10.0<p\leq30.0$,
$30.0<p\leq100.0$, $p>100.0$, where $p$ is in atm. ChemGen utilizes
an optimization where $\ln p$ is pre-calculated before traversing
over the booleans, where its value is checked against the pre-calculated
$\ln p_{i}$ values.

\section{Time integration strategies}

The stability of the CFD time integration may be compromised by explicitly
applying the source term of Equation~(\ref{enu:source}). Different
approaches can be utilized by time-splitting the evolution of the
chemical state and retain the stability of the fluid dynamic algorithms~\citep{Str68,Hou11,Lv15,Joh20}.
These approaches typically do not involve the flux from either convection
or diffusion. By removing any spatial fluxes, the conservation Equation~(\ref{eq:conservation-law-strong-form})
reduces to an ordinary differential equation,

\begin{equation}
\frac{dy_{f}}{dt}=\mathcal{S}\left(y_{f}\right).\label{eq:conservation-law-to-ode}
\end{equation}
By removing the spatial fluxes, the fluid state from Equation~(\ref{eq:reacting-navier-stokes-state})
is reduced to

\begin{equation}
y_{f}=\left(0,\ldots,0,\rho u,C_{1},\ldots,C_{n_{s}}\right).\label{eq:ODE_state}
\end{equation}
By inspecting the source term,

\begin{equation}
\mathcal{S}\left(y_{f}\right)=\left(0,\ldots,0,0,\omega_{1},\ldots,\omega_{n_{s}}\right)^ {},\label{eq:RNS_source}
\end{equation}
and analyzing the time rate of change of the internal energy,

\begin{equation}
\frac{\partial\rho u}{\partial t}=\frac{\partial\rho u}{\partial T}\frac{\partial T}{\partial t}+\sum_{i=1}^{n_{s}}\frac{\partial\rho u}{\partial C_{i}}\frac{\partial C_{i}}{\partial t}=0,\label{eq:derive_omega_T}
\end{equation}
the temperature source term is realized as

\begin{equation}
\frac{\partial T}{\partial t}=\omega_{T}=-\frac{\sum_{i=1}^{n_{s}}W_{i}u_{i}\omega_{i}}{c_{v}}.\label{eq:temperature_source}
\end{equation}
This yields the chemical evolution system solved by ChemGen

\begin{equation}
\begin{array}{ccc}
\frac{dy_{c}}{dt}=S\left(y_{c}\right),\quad & y_{c}=\left(T,C_{1},\ldots,C_{n_{s}}\right),\quad & S\left(y_{c}\right)=\left(\omega_{T},\omega_{1},\ldots,\omega_{n_{s}}\right).\end{array}\label{eq:extracted_problem-1}
\end{equation}
For brevity we now refer to $y$ as the chemicals state, $y=y_{c}$.
In this section we will cover some techniques implemented in ChemGen
to integrate this state in time. 

The formulation presented in Equation~(\ref{eq:extracted_problem-1}).
is derived using a constant internal energy assumption and thus allows
for pressure to change. Some systems require a constant pressure assumption.
Appendix~(\ref{subsec:Constant-Pressure-Strategy}) covers how a
constant pressure formulation can be extracted using an assumption
that uses a global $\rho h$.

\subsection{Explicit Runge-Kutta time integration method}

ChemGen provides a Runge-Kutta 4th-order method (RK4) for integrating
the chemical state from $y^{n}$ to $y^{n+1}$ over a time interval
of $\Delta t$ from Equation~(\ref{eq:extracted_problem-1}). This
explicit time integration scheme approximates the solution by evaluating
the source term at four intermediate stages within each time step~\citep{Hai96}.
Given state $y_{n}$ at time $t_{n}$, the RK4 method updates to the
next state, $y^{n+1}$ at time $t^{n+1}$. The algorithm is

\begin{equation}
\begin{array}{ccc}
\Gamma_{1} & = & S\left(y^{n}\right)\\
\Gamma_{2} & = & S\left(y^{n}+\frac{\Delta t}{2}\Gamma_{1}\right)\\
\Gamma_{3} & = & S\left(y^{n}+\frac{\Delta t}{2}\Gamma_{2}\right)\\
\Gamma_{4} & = & S\left(y^{n}+\Delta t\Gamma_{3}\right)\\
y^{n+1} & = & y^{n}+\frac{\Delta t}{6}\left(\Gamma_{1}+2\Gamma_{2}+2\Gamma_{3}+\Gamma_{4}\right)
\end{array}\label{eq:RK4}
\end{equation}

This method achieves fourth-order accuracy in time, meaning the local
truncation error scales as $\mathcal{O}(\Delta t^{5})$ and the global
error as $\mathcal{O}(\Delta t^{4})$. RK4 is effective for non-stiff
problems where high accuracy is desired without the overhead of the
implicit solvers, which, as detailed below, require potentially expensive
linear solvers.

\subsection{Implicit time integration and Jacobians\label{subsec:Implicit-Time-Integration}}

Although explicit methods like RK4 offer high accuracy, they can suffer
from stability issues, particularly for stiff problems. In contrast,
implicit time integration methods provide improved stability at the
cost of increased computational overhead. Consider first the backward
Euler implicit time integration method, which seeks $y^{n+1}$ that
satisfies

\begin{equation}
\frac{y^{n+1}-y^{n}}{\Delta t}=S\left(y^{n+1}\right).\label{eq:Backward_euler}
\end{equation}
 Newton's method can be used to find $y^{n+1}$ via iteration by solving

\begin{equation}
\mathcal{G}\left(y_{k}^{n+1}\right)\left(y_{k+1}^{n+1}-y_{k}^{n+1}\right)=-f(y_{k}^{n+1}),\quad f(y_{k}^{n+1})=\frac{y^{n+1}_k-y^{n}}{\Delta t}-S\left(y_k^{n+1}\right),\label{eq:newtons-method-2}
\end{equation}
for $y_{k+1}^{n+1}$, the state at the $\left(k+1\right)$th Newton
iteration. $\mathcal{G}\left(y_{k}^{n+1}\right)$ is the Jacobian
for the time integration given by

\begin{equation}
\mathcal{G}\left(y_{k}^{n+1}\right)_{ij}=\frac{\partial f(y_{k,i}^{n+1})}{\partial y_{j}}=\mathcal{D}_{ij}-\mathcal{J}_{ij},\quad\mathcal{J}_{ij}=\frac{\partial S_{i}\left(y_{k}^{n+1}\right)}{\partial y_{j}},\label{eq:jacobian}
\end{equation}
where $\mathcal{D}_{ij}$ depends on the time integration scheme ($\mathcal{D}_{ij}=\delta_{ij}/\Delta t$
for the backward Euler method) and $\mathcal{J}_{ij}$ is the Jacobian
of the chemical source term with $y_{j}$ as the $j^{th}$ element of
$y$ and $S_{i}$ is the $i^{th}$ element of of $S$.

ChemGen employs analytical differentiation to evaluate the chemical
source term Jacobian, $\mathcal{J}_{ij}$. The code generation is
optimized to compute only necessary components for efficient matrix
assimilation. The elements of the species concentration source term
Jacobian are

\begin{equation}
\frac{\partial\omega_{k}\left(y\right)}{\partial C_{j}}=\frac{\partial\omega_{k}}{\partial C_{j}}+\frac{\partial\omega_{k}}{\partial T}\frac{\partial T}{\partial C_{j}},\label{eq:derivative_matrices}
\end{equation}
where
\begin{equation}
\frac{\partial T}{\partial C_{j}}=-\left(\frac{\partial\rho u}{\partial T}\right)^{-1}\frac{\partial\rho u}{\partial C_{j}}=-\frac{1}{\rho c_{v}}W_{j}u_{j}=-\frac{W_{j}u_{j}}{\rho c_{v}}.\label{eq:temperature_wrt_species}
\end{equation}
With these derivatives known an $n_{s}+1$ by $n_{s}+1$ matrix is
formed,

\begin{equation}
\mathcal{J}_{ij}=\begin{cases}
\frac{\partial\omega_{T}}{\partial T} & i=1;\quad j=1\\
\frac{\partial\omega_{T}}{\partial C_{j-1}} & i=1;\quad j>1\\
\frac{\partial\omega_{i-1}}{\partial T} & i>1;\quad j=1\\
\frac{\partial\omega_{i-1}}{\partial C_{j-1}} & i>1;j>1
\end{cases},\label{eq:derivative_matrices_temp_evol}
\end{equation}
where

\begin{equation}
\frac{\partial\omega_{T}}{\partial T}=-\frac{\sum_{i=1}^{n_{s}}W_{i}\frac{\partial u_{i}}{\partial T}\omega_{i}}{c_{v}}+\frac{\sum_{i=1}^{n_{s}}W_{i}\frac{\partial u_{i}}{\partial T}\omega_{i}}{c_{v}^{2}}\frac{\partial c_{v}}{\partial T}\label{eq:temperature_source_2-1}
\end{equation}
and

\begin{equation}
\frac{\partial\omega_{T}}{\partial C_{j-1}}=-\frac{\sum_{i=1}^{n_{s}}W_{i}\frac{\partial u_{i}}{\partial T}\frac{\partial\omega_{i}}{\partial C_{j-1}}}{c_{v}}+\frac{\sum_{i=1}^{n_{s}}W_{i}u_{i}\omega_{i}}{c_{v}^{2}}\frac{\partial c_{v}}{\partial C_{j-1}}.\label{eq:temperature_source_2-1-1}
\end{equation}

\subsection{Additional time integrators}

While the backward Euler time integration is inherently stable, it
may exhibit poor accuracy or convergence for certain chemical models
and time step sizes. Therefore, ChemGen offers additional implicit
integrators, as listed in Table~\ref{tab:Time_integrators} with
formulation details in Appendix~(\ref{sec:Algorithms-of-Implicit}).
The appropriate choice of time integrator is problem dependent. For
example, in low Mach number solvers where acoustic time scales are
removed, chemical integration may proceed on the order of microseconds.
In contrast, fully compressible solvers may require time steps as
small as nanoseconds or even less, in which case a lower-order but
computationally faster time integrator may be more suitable.

\begin{table}[H]
\caption{Implicit time integration strategies provided by ChemGen. $n_{\mathrm{max}}$
is the maximum number of Newton iterations.\label{tab:Time_integrators}}
\begin{tabular}{|c|c|c|c|c|}
\hline 
Strategy & Order & Stages & \# Linear Solves & Section/Appendix\tabularnewline
\hline 
\hline 
SDIRK-2 & 2nd & 2 & up to $n_{\mathrm{max}}$ per stage & \ref{subsec:SDIRK-2}\tabularnewline
\hline 
SDIRK-4 & 4th & 5 & up to $n_{\mathrm{max}}$ per stage & \ref{subsec:SDIRK-4}\tabularnewline
\hline 
Rosenbrock & 2nd & 2 & one per stage & \ref{subsec:Rosenbroc}\tabularnewline
\hline 
YASS & 1st & 1 & one & \ref{subsec:YASS}\tabularnewline
\hline 
Backward Euler & 1st & 1 & up to $n_{\mathrm{max}}$ & Section \ref{subsec:Implicit-Time-Integration}\tabularnewline
\hline 
\end{tabular}
\end{table}

Other time integration strategies not yet implemented in ChemGen can
be added using the same framework as the existing methods. These include,
but are not limited to, multi-step implicit methods such as backward
differentiation formulas, which utilize information from previous
time steps, and higher-order Rosenbrock methods ~\citep{Zha14},
which require additional stages. Integration methods such as DGODE
~\citep{Joh20} and CHEMEQ2~\citep{Mot01} may be more challenging
to implement. DGODE is based on finite element numerics, while CHEMEQ2
requires the source term to be separated into distinct production
and consumption components, introducing additional complexities not
currently addressed by ChemGen.

\subsection{Linear solvers}

The aforementioned implicit time integrators require the solution
to the linear system~(\ref{eq:newtons-method-2}), recast here as

\begin{equation}
Ax=b.\label{eq:matrix_example_1-1}
\end{equation}
It is well-established that employing a direct solver requires $\mathcal{O}\left(n^{3}\right)$
operations for an $n\times n$ square matrix~\citep{Gol13}. Although ChemGen supplies a direct solver, its default linear solver is a stand alone function that uses the Generalized Minimal Residual (GMRES) method in order to mitigate the $\mathcal{O}(n^{3})$ cost~\citep{Saa86}.
 GMRES is an iterative Krylov subspace method that builds
an orthonormal basis of the Krylov subspace and minimizes the residual
over this subspace at each iteration. 

The number of GMRES iterations can be reduced by applying a preconditioner,
$M$, as 
\begin{equation}
M^{-1}Ax=M^{-1}b.\label{eq:matrix_example_preconditioner}
\end{equation}
ChemGen provides two standard options: Jacobi preconditioning and
Gauss-Seidel preconditioning. In the former, $M$ simply consists
of the diagonal of $A$, which can be efficiently inverted by computing
the reciprocal of each diagonal entry. In the latter, $M$ consists
of the lower triangular portion of $A$ (including the diagonal),
which can be easily applied with forward substitution. For more advanced
preconditioning techniques used in chemical kinetics, see, for example,
~\citep{Wal23}.

ChemGen is not intended to serve as a comprehensive linear-solver
library. For more advanced or scalable linear solver capabilities,
ChemGen can be interfaced with external libraries such as Eigen~\citep{Gue10}
or PETSc~\citep{Bal97}, which offer efficient preconditioned Krylov
solvers and direct sparse solvers for large systems. Additionally,
while ChemGen includes several implicit time integration strategies
(e.g., backward Euler, SDIRK methods), offloading the integration
of stiff ODE systems to established libraries like CVODES~\citep{Ser05}
is also a reasonable approach. Such packages allow user-supplied source
terms and Jacobians that can then be used in time integration strategies.

\section{Code generation\label{sec:Code-Generation}}

\subsection{The concept of decorators}

\Cpp is a malleable language that allows developers to abstract underlying
data structures through type aliases (\texttt{using scalar = float;})
and operator overloading. As a result, variables may appear to be
simple scalars while in fact representing complex types with custom
behavior. This flexibility enables expressive, domain-specific syntax,
but this means that there is no single set of generated code that
can be embedded into all existing \Cpp software.

For instance, it may be more suitable for the function

\begin{lstlisting}[language=C++, numbers=left, numberstyle=\tiny, stepnumber=1,
  frame=single, breaklines=true, basicstyle=\ttfamily\small]
float sqr (float a){return a * a;} \\example function 1
\end{lstlisting}

to be declared as a \texttt{const} in a \texttt{struct} or \texttt{class}.
In addition, the scalar \texttt{float} could also be \texttt{double},
and the user may wish to pass the function parameters by \texttt{const}
reference, i.e., \texttt{float a} $\rightarrow$ \texttt{const float\& a}.
A function annotation may also be desirable, such as a preprocessor
macro provided the Kokkos library~\citep{Tro22}. This can turn the
above example to

\begin{lstlisting}[language=C++, numbers=left, numberstyle=\tiny, stepnumber=1,
  frame=single, breaklines=true, basicstyle=\ttfamily\small]
KOKKOS_INLINE_FUNCTION
float sqr (const float& a) const {return a * a;} \\example function 2
\end{lstlisting}

ChemGen utilizes \emph{decorators} to assist in the flexibility of
generating desirable \Cpp code. The above example can be changed
to a Python formattable Python string as

\begin{lstlisting}[language=Python, frame=single, breaklines=true,
  basicstyle=\ttfamily\small, numbers=none]
{device_function}
{scalar_function} sqr ({scalar_parameter} a) {const_option} {{return a * a;}} \\Python formattable
\end{lstlisting}

ChemGen utilizes these decorators throughout the code generation process.
When the code generation takes place, a \texttt{configuration.yaml}
file is utilized that replaces the decorators with the appropriate
text. So for example function 1 is

\begin{lstlisting}[language=YAML, numbers=left, numberstyle=\tiny, stepnumber=1,
  frame=single, breaklines=true, basicstyle=\ttfamily\small]
decorators:
  scalar_function: "float"
  scalar_parameter: "float"
  device_option: ""
  const_option: ""
\end{lstlisting}

and for example function 2 is

\begin{lstlisting}[language=YAML, numbers=left, numberstyle=\tiny, stepnumber=1,
  frame=single, breaklines=true, basicstyle=\ttfamily\small]
decorators:
  scalar_function: "float"
  scalar_parameter: "const float&"
  device_option: "KOKKOS_INLINE_FUNCTION"
  const_option: "const"
\end{lstlisting}

The configuration file also includes the ability to switch from \texttt{std::array}
to \texttt{std::vector} and allows other data types such as \texttt{Views}
from Kokkos or \texttt{std::tuple}-derived types.

\subsection{Derivatives}

ChemGen employs a mechanical differentiation approach. It supplies
a differential codebase that contains derivatives of common functions
and their chain rules. For instance, the Arrhenius function as generated
using the standard configuration in ChemGen is

\begin{lstlisting}[language=C++, numbers=left, numberstyle=\tiny, stepnumber=1,
  frame=single, breaklines=true, basicstyle=\ttfamily\small]
double
arrhenius(const double& A, const double& B, const double& E, const double& temperature) 
{
        double pow_term = pow_gen(temperature, B);
        double exp_term = exp_gen(divide(-E, universal_gas_constant() * temperature));

        return multiply(A,
                        multiply(pow_term,
                                 exp_term));
}
\end{lstlisting}

ChemGen supplies the derivatives for the mathematical operations contained
in the \texttt{arrhenius} function. For example, the derivatives
of \texttt{pow\_gen}, given that it has two parameters, would yield
two derivatives and one chain rule,

\begin{lstlisting}[language=C++, numbers=left, numberstyle=\tiny, stepnumber=1,
  frame=single, breaklines=true, basicstyle=\ttfamily\small]
double dpow_da(const double& a, const double& b)  
{return b * pow_gen(a, b - double(1));}
double dpow_db(const double& a, const double& b)  
{return pow_gen(a, b) * log_gen(a);}

double
pow_gen_chain(const double& a, 
              const double& a_perturbation,
              const double& b,
              const double& b_perturbation) 
{ 
    return dpow_da(a, b) * a_perturbation + dpow_db(a, b) * b_perturbation;
}
\end{lstlisting}

For all existing lower-level functions, the partial derivatives with
respect to each parameter, as well as the associated chain rule contractions
with perturbations, can be assumed accessible when developing higher-level
functions in ChemGen. With this capability, the derivative of the
Arrhenius function with respect to temperature is immediately available,

\begin{lstlisting}[language=C++, numbers=left, numberstyle=\tiny, stepnumber=1,
  frame=single, breaklines=true, basicstyle=\ttfamily\small]
double
darrhenius_dtemperature(const double& A, 
                        const double& B,
                        const double& E,
                        const double& temperature) 
{
        double pow_term = pow_gen(temperature, B);
        double dpow_term_dtemperature = dpow_da(temperature, B);

        double exp_term =
            exp_gen(divide(-E,
                           universal_gas_constant() * temperature));
        double dexp_term_dtemperature =
            exp_chain(divide(-E,
                             universal_gas_constant() * temperature),
                      ddivide_db(-E,
                                 universal_gas_constant() * temperature));

        return multiply(A,
                        multiply_chain(pow_term,
                                       dpow_term_dtemperature,
                                       exp_term,
                                       dexp_term_dtemperature));
}
\end{lstlisting}

The mechanical differentiation approach is both efficient and more
readable than symbolic differentiation using tools such as SymPy~\citep{sympy}.
While symbolic differentiation can be valuable, the mechanical approach
is better suited for organizing and maintaining ChemGen’s code generation
process. Regardless, since the functions are available in \Cpp, packages
such as autodiff~\citep{Lea18} can be implemented to use a wider
variety of differentiable functionality. For a mathematical overview
of the derivatives needed for chemically reacting flow, see~\citep{Nie17}.

\subsection{Generation process}

In the ChemGen software, there are three main file types. The file
type consists of Python files that manipulate chemistry data and existing
static code to create compilation-ready source code. The second file
type comprises static code files that are ingested by the aforementioned
Python code and reformatted to the desired target software using decorators.
The final type is a mixture of Python code used in tutorials, most
of which are covered in Section~(\ref{sec:Results-and-Demonstration}),
and \texttt{yaml} files that contain the chemical model data.

\begin{figure}[H]
\begin{centering}
\includegraphics{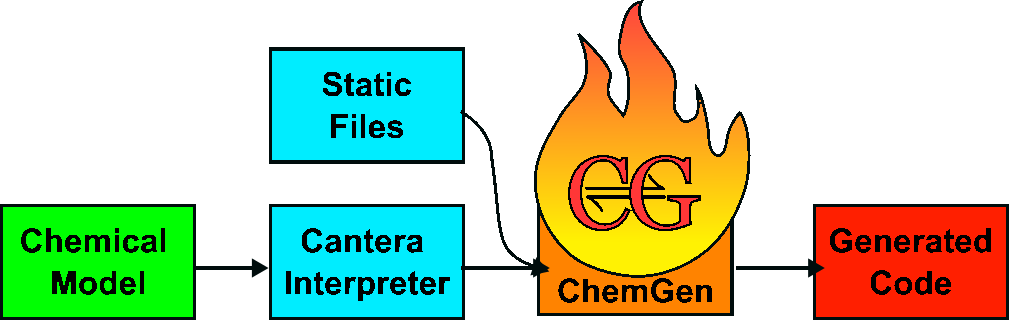}
\par\end{centering}
\caption{\label{fig:flowchart}ChemGen flowchart}
\end{figure}

The code generation process in ChemGen is designed to be as streamlined
as possible and is depicted in Figure~\ref{fig:flowchart}. ChemGen
relies on Cantera to interpret chemical model files. Cantera performs
several key tasks: parsing the chemistry data, converting units to
the International System of Units (SI), and providing thermodynamic
relationships that are refit into the desired polynomial form. As
a well-maintained and widely used tool in the combustion community,
Cantera provides a dependable foundation for ChemGen. Nevertheless,
ChemGen could also be extended to handle data parsing, unit conversion,
and thermodynamic relationships for refitting internally.

After the Cantera step, ChemGen begins to generate source code. The
state size is hardcoded with several automatically generated variables,
\texttt{Species}, \texttt{ChemicalState}, \texttt{Jacobian} . The
stoichiometric coefficients ($\nu_{ij}$, $\nu_{ij}^{f}$, and $\nu_{ij}^{r}$)
and reaction coefficients, which depend on the reaction type, are
then accrued. Each reaction rate is generated into a\texttt{reactions.h}
header file; for example, if the first reaction of a Cantera-based
yaml file is

\begin{lstlisting}[language=YAML, numbers=left, numberstyle=\tiny, stepnumber=1,
  frame=single, breaklines=true, basicstyle=\ttfamily\small]
- equation: H + O2 <=> O + OH
rate-constant: {A: 1.0399e+11, b: 0.0, Ea: 6.405704e+07}
\end{lstlisting}

then the following code is generated with standard configuration settings,
\begin{lstlisting}[language=C++, numbers=left, numberstyle=\tiny, stepnumber=1,
  frame=single, breaklines=true, basicstyle=\ttfamily\small]
double call_forward_reaction_0(const double& temperature, 
                              const double& log_temperature) 
{ 
	return arrhenius(103990000000.0, 64057040.0, temperature, log_temperature);
}
\end{lstlisting}

\noindent the units here are in SI. Once all reactions are generated, a chemical
source term file is created. The source term file begins with generated
code that calculates $M_{c}$, $\ln T$, and $p$, as well as $\Delta G_{j}'$
for all reactions, \texttt{gibbs\_reactions}. ChemGen then loops
through all reactions, and generates the forward rates and the corresponding
equilibrium constants for reversible reactions. In the above example,
a reverse reaction is necessary, so ChemGen generates

\begin{lstlisting}[language=C++, numbers=left, numberstyle=\tiny, stepnumber=1,
  frame=single, breaklines=true, basicstyle=\ttfamily\small]
double forward_reaction_0 = call_forward_reaction_0(temperature, log_temperature);
double equilibrium_constant_0 = exp_gen(-gibbs_reactions[0]);
double rate_of_progress_0 = 
species[0] * species[3] * forward_reaction_0 
- species[2] * species[4] * forward_reaction_0/equilibrium_constant_0;
\end{lstlisting}

\noindent While many options can be specified during the generation process,
the simplest approach requires only the following terminal command
to generate source code using ChemGen:

\begin{lstlisting}[language=bash, frame=single, breaklines=true,
  basicstyle=\ttfamily\small, numbers=none]
chemgen.py mech.yaml .
\end{lstlisting}

\noindent This command, as long as \texttt{mech.yaml} is a Cantera supported
file, generates \Cpp code in the current directory in a \texttt{src}
folder using the standard settings.

\section{Results and demonstration cases\label{sec:Results-and-Demonstration}}

This section presents standalone test cases to confirm that the generated
source code is behaving as intended. We then embed the generated code
in detonationEulerFoam~\citep{Sun23} based on the OpenFOAM package~\citep{Ope25}
to demonstrate its performance in a CFD simulation of a moving detonation
wave.

\subsection{Source term accuracy\label{subsec:Source-Term-Accuracy}}

To test the source term accuracy, we randomly generate a chemical
state as

\begin{eqnarray}
T & = & 1000+r_{T}1500\text{ K}\nonumber \\
p & = & 0.1+9.9r_{p}\text{ atm}\nonumber \\
X_{i} & = & r_{i},\quad\text{for }i=1,\dots,n_{s},\label{eq:chemical_state_01}
\end{eqnarray}
where are random values $r_{T},r_{p},r_{1},\dots,r_{n_{s}}$ are generated from a uniform distribution, $\sim\mathcal{U}(0,1)$. $X_{i}$ is the mole fraction of the
$i^{th}$ species, defined as

\begin{equation}
X_{i}=\frac{C_{i}}{\sum_{i=1}^{n_{s}}C_{i}},\label{eq:mole_fraction}
\end{equation}
which is renormalized after all $n_{s}$ species are randomly generated.
We then calculate the concentrations as $C_{i}=X_{i}p/\left(RT\right)$,
forming the complete chemical state for ChemGen, $y_{c}$. We consider
five different chemical models with increasing size that are common
to simulations of chemically reacting flows and summarized in Table
~\ref{tab:chemical models}. For this analysis we generated 10,000
random chemical states generated for each model.

\begin{table}[H]
\begin{centering}
\begin{tabular}{|c|c|c|c|}
\hline 
Model & $n_{s}$ & $n_{r}$ & Reference\tabularnewline
\hline 
\hline 
Ó Connaire  & 10 & 40 & \citep{Oco04} \tabularnewline
\hline 
Burke & 13 & 27 & \citep{Bur12} \tabularnewline
\hline 
GRI-Mech 3.0  & 53 & 325 & \citep{Smi00}\tabularnewline
\hline 
UCSD  & 63 & 268 & \citep{SanDiegoMech} \tabularnewline
\hline 
FFCM-2  & 96 & 1054 & \citep{Zha23} \tabularnewline
\hline 
\end{tabular}
\par\end{centering}
\centering{}\caption{Summary of chemical models analyzed \label{tab:chemical models}}
\end{table}

We use Cantera's source term prediction, \texttt{net\_production\_rates}
, for comparison in this analysis. The average relative error per
chemical state is

\begin{equation}
\epsilon=\frac{1}{n_{s}}\sum_{i=1}^{n_{s}}\left(\begin{cases}
\left|\frac{S_{cg,i}-S_{ct,i}}{S_{ct,i}}\right| & \text{if }\left|S_{ct,i}\right|>10^{-10}\\
0 & \text{otherwise}
\end{cases}\right),\label{eq:source_error}
\end{equation}
where $S_{\mathrm{ct},i}$ denotes the Cantera-predicted source term
for the $i^{th}$ specie and $S_{\mathrm{cg},i}$ denotes the corresponding
ChemGen-predicted source term. If the magnitude of the Cantera source
term is less than $10^{-10}$, the error is considered to be zero.

\begin{figure}[H]
\begin{centering}
\includegraphics[scale=0.8]{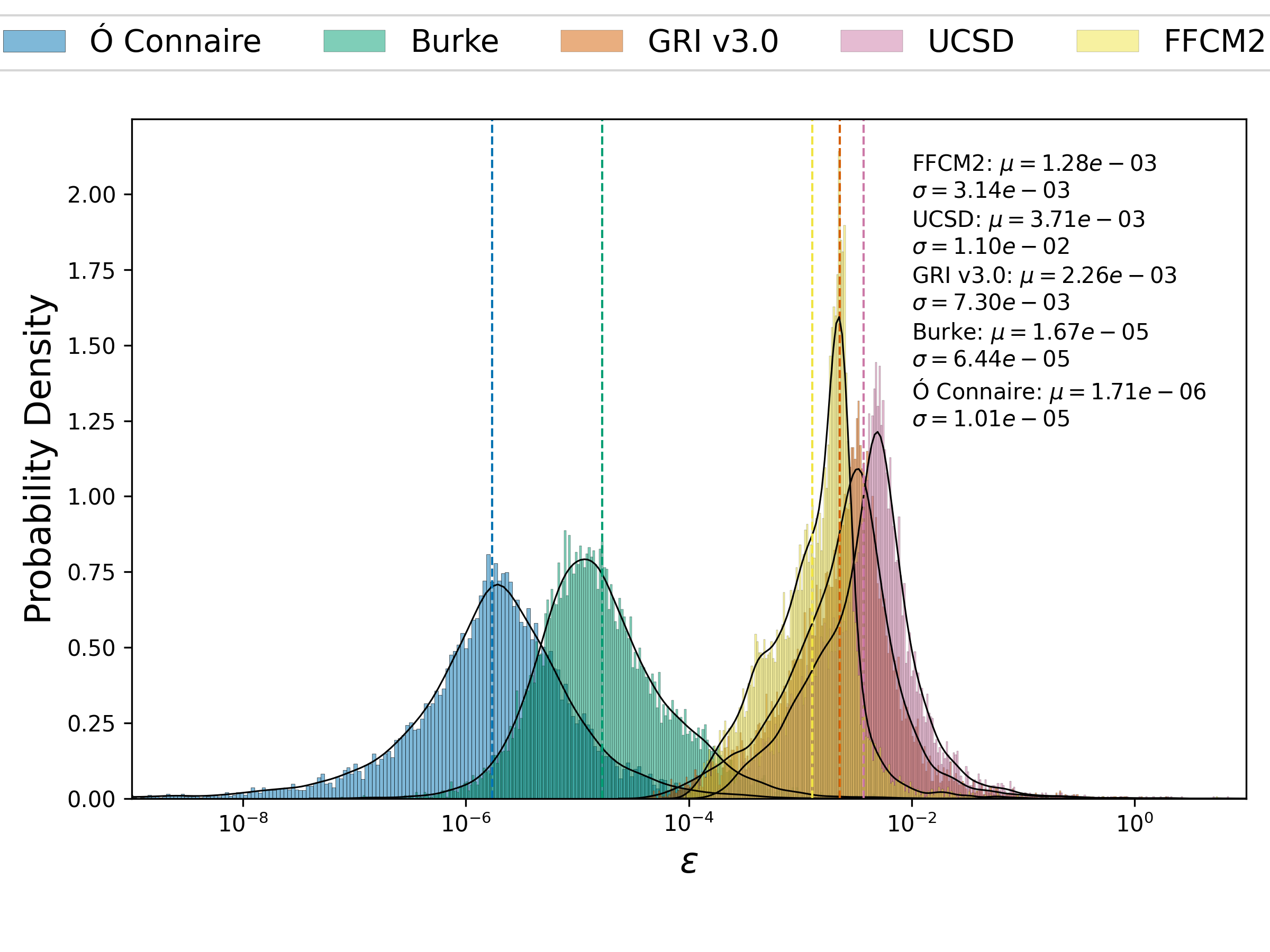}
\par\end{centering}
\caption{\label{fig:source_term_accuracy}Distribution of error, $\epsilon$,
as defined in Equation~(\ref{eq:source_error}), for ChemGen source
calculations in comparison to Cantera source calculations for five
tested chemical models from Table~\ref{tab:chemical models}. The
vertical dash line represents the mean of the distribution and is
colored according to the corresponding model. The distribution mean,
$\mu$, and corresponding standard deviation, $\sigma$, are also
reported as text. }
\end{figure}

The distribution of errors for the randomly generated states and the
chemical models analyzed are shown in Figure~(\ref{fig:source_term_accuracy}).
Agreement with Cantera predictions varies with model complexity. Among
the models tested, the UCSD model exhibits the largest discrepancy.
While the ChemGen implementation is designed to be consistent with
Cantera’s formulation, differences arise primarily due to the thermodynamic
polynomial representations. Cantera supports multiple polynomial formats
for thermodynamic property calculations, whereas ChemGen refits all
species thermodynamic properties to a user-specified polynomial order
during code generation. The ChemGen refits for the data in Figure~(\ref{fig:source_term_accuracy})
are based on $n_{p}=7$. To investigate the effect of polynomial degree,
we repeat the error analysis with different choices of $n_{p}$ which
confirms an inverse relationship between the mean relative error and
the polynomial degree. The details from this further analysis are
shown in \ref{subsec:sourcetermpoly}.

\subsection{Explicit time integration\label{subsec:Explicit-Time-Integration}}

\begin{figure}[H]
\begin{centering}
\includegraphics[scale=0.55]{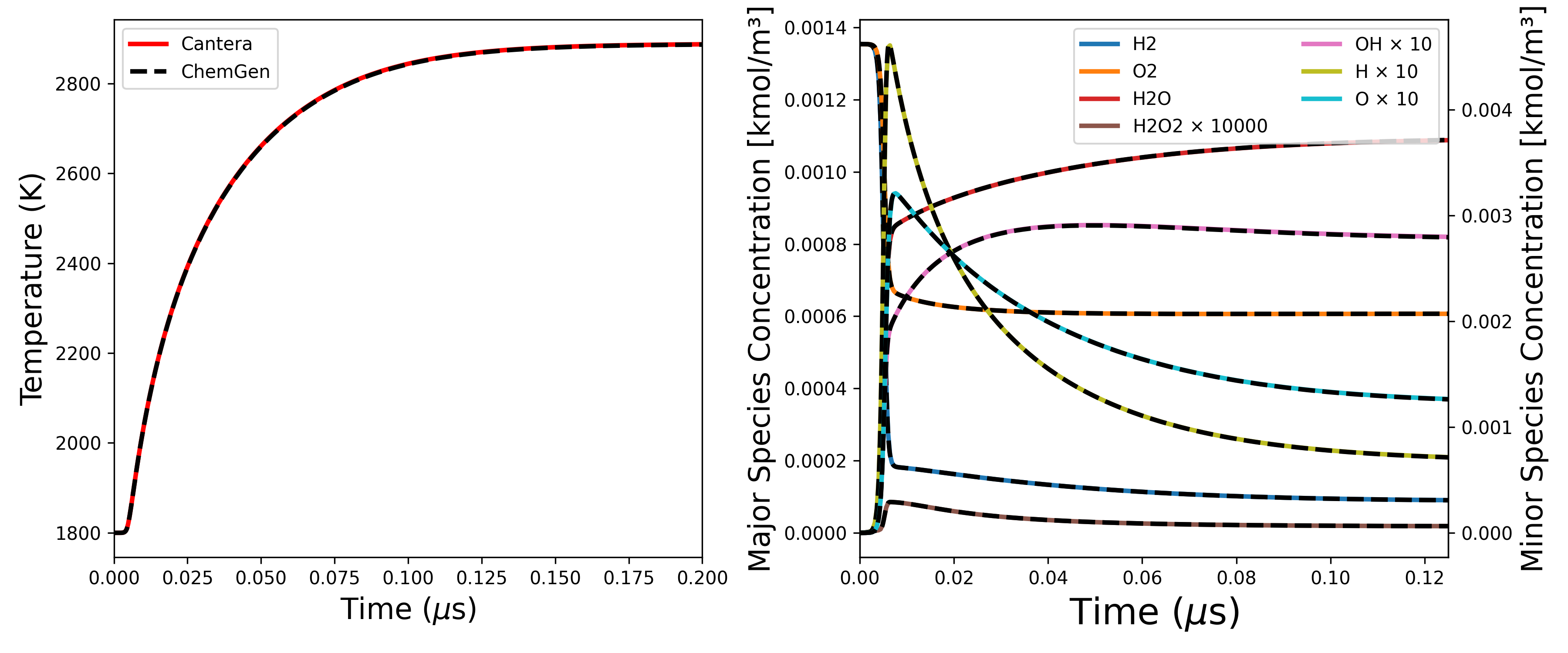}
\par\end{centering}
\caption{\label{fig:RK4_T_C}Temperature (left) and species (right) evolution
for a constant volume reactor using an explicit RK4 scheme and initial
conditions given in Equation~(\ref{eq:chemical_state_02}). These
results are produced using a hydrogen submodel extracted from FFCM2
\citep{Zha23}. The solid lines represent the Cantera solution, and
the dashed black lines represent the ChemGen solution.}
\end{figure}

With confidence in the source term calculation, we proceed to apply
ChemGen's explicit RK4 time integrator to temporally evolve the following
initial chemical state:

\begin{eqnarray}
T & = & 1800\nonumber \\
p & = & 101325.0\nonumber \\
\left(X_{H_{2}},X_{O_{2}},X_{N_{2}}\right) & = & \left(0.2,0.2,0.6\right).\label{eq:chemical_state_02}
\end{eqnarray}
We use a hydrogen submodel extracted from FFCM-2~\citep{Zha23},
which is created by removing carbon containing species, electronically
excited states, such as $OH^{*}$, and inert species $Ar$ and $He$.
The resulting submodel consists of 9 species and 25 reactions. We
calculate the concentrations, $C_{i}$, in the same manner as in Section~(\ref{subsec:Source-Term-Accuracy}).
We employ a time-step size of $\Delta t=1\times10^{-8}$ seconds.
With larger time-step sizes, the RK4 time integrator becomes unstable.
The resulting temperature and species concentration profiles are shown
in Figure~(\ref{fig:RK4_T_C}) , where the ChemGen results are represented
by dashed black lines and Cantera results are represented by solid
colored lines. We produce the Cantera results using the ReactorNet
class.

Following a brief induction period, the chemical state undergoes a
sudden temperature rise accompanied by rapid species changes, where
the reactants, $H_{2}$ and $O_{2}$, are consumed and the product,
$H_{2}O$, along with intermediate species, $H_{2}O_{2}$, $OH$, $O$,
and $H$, are created. This is followed by a more gradual evolution
toward a steady state. The main reactants and products are referred
to as major species, which for this model are $H_{2}$, $O_{2}$,
and $H_{2}O$. The intermediate and small remnant species are referred
to as minor species and usually exist in trace quantities ($X_{i}<1\times10^{-3}$).
Therefore, scaling multipliers are applied to minor species in Figure~(\ref{fig:RK4_T_C})
in order to easily visualize their profiles on the same graph.

\subsection{Jacobian verification\label{subsec:Jacobian-Verification}}

Next, we verify the Jacobian calculations. Recall the Jacobian supplied
by ChemGen is

\begin{equation}
\mathcal{J}_{ij}=\begin{cases}
\frac{\partial\omega_{T}}{\partial T} & j=1\\
\frac{\partial\omega_{T}}{\partial C_{j-1}} & i=1;\quad j>1\\
\frac{\partial\omega_{i-1}}{\partial T} & i>1;\quad j=1\\
\frac{\partial\omega_{i-1}}{\partial C_{j-1}} & i>1;j>1
\end{cases}.\label{eq:Ja_Jb_results}
\end{equation}
In this section we present verification of the Jacobians using the
finite difference method. To calculate the finite difference approximation
of the Jacobians, we utilize the second order finite difference approximation
for each element by perturbing each species forward,

\begin{equation}
C_{j}^{+}=\left(C_{1}+\delta_{1j}\delta C,\dots,C_{n_{s}}+\delta_{n_{s}j}\delta C\right),\label{eq:Ja_Jb_results-1}
\end{equation}
and backwards, 

\begin{equation}
C_{j}^{-}=\left(C_{1}-\delta_{1j}\delta C,\dots,C_{n_{s}}-\delta_{n_{s}j}\delta C\right).\label{eq:Ja_Jb_results-1-1}
\end{equation}
These perturbations are utilized in the finite difference Jacobian,

\begin{eqnarray}
\hat{\mathcal{J}_{ij}} & = & \begin{cases}
\frac{\omega_{T}(C,T+\delta T)-\omega_{T}(C,T-\delta T)}{2\delta T} & i=1;\quad j=1\\
\frac{\omega_{T}(C_{j-1}^{+},T)-\omega_{T}(C_{j-1}^{-},T)}{2\delta C} & i=1;\quad j>1\\
\frac{\omega_{i-1}(C,T+\delta T)-\omega_{i-1}(C,T-\delta T)}{2\delta T} & i>1;\quad j=1\\
\frac{\omega_{i-1}\left(C_{j-1}^{+},T\right)-\omega_{i-1}\left(C_{j-1}^{-},T\right)}{2\delta C} & i>1;\quad j>1,
\end{cases}\label{eq:finite_difference_Tevolve}
\end{eqnarray}
where $\delta_{ij}$ is from Equation~(\ref{eq:jacobian}), $\delta T$
and $\delta C$ are the temperature and concentration perturbations,
respectively. An initial test to verify the Jacobian
is to check the order of convergence of the finite difference errors
as compared to the analytical Jacobian provided by ChemGen.

\begin{figure}[H]
\begin{centering}
\includegraphics[scale=0.8]{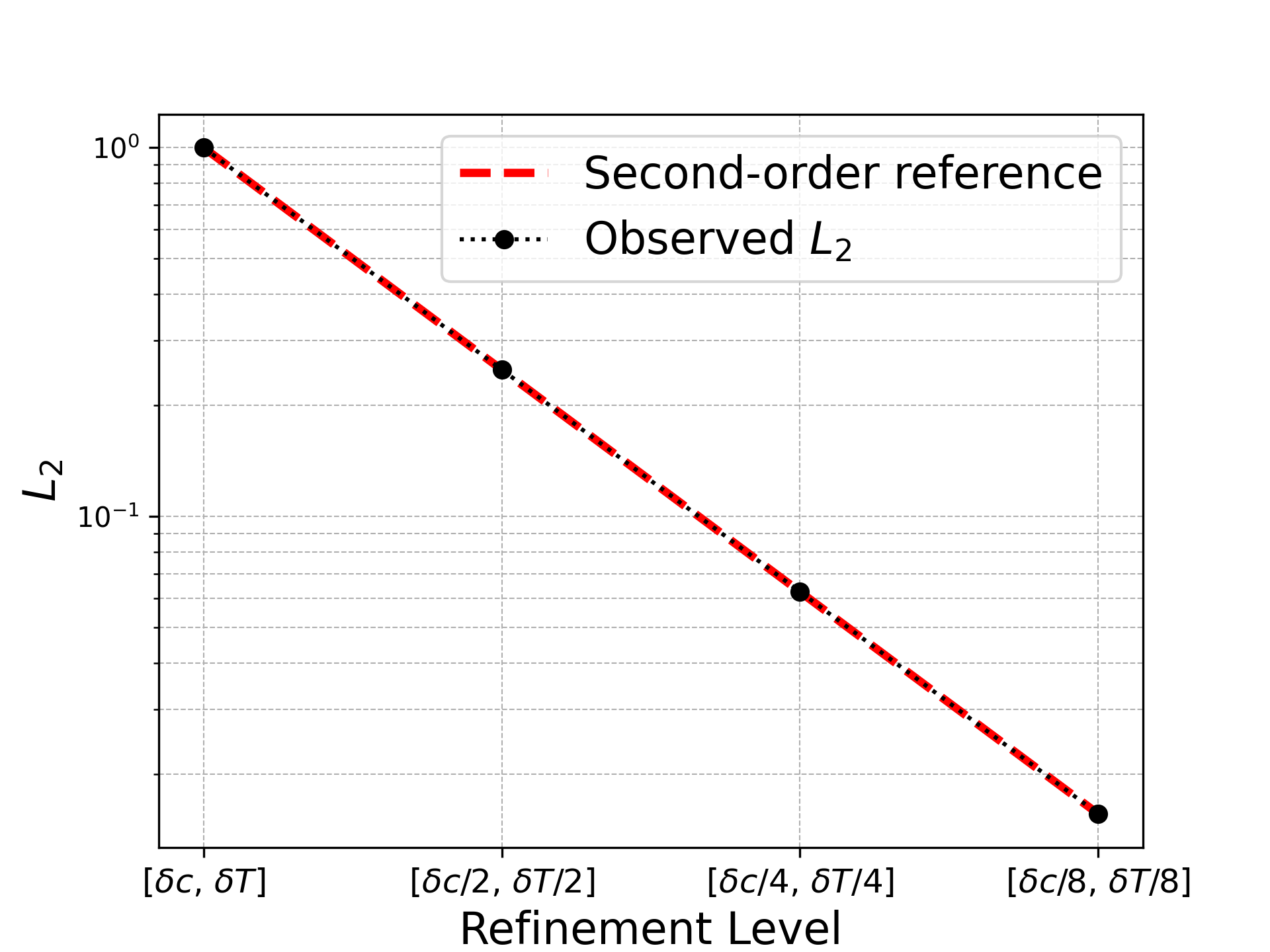}
\par\end{centering}
\caption{\label{fig:OOA}Order of accuracy convergence for $\hat{\mathcal{J}_{ij}}$
when compared to $\mathcal{J}_{ij}$.}
\end{figure}

We use the $L_{2}$-norm to assess the error of the finite difference
approximation,

\begin{equation}
L_{2}=||\mathcal{J}_{ij}-\hat{\mathcal{J}_{ij}}||=\sqrt{\sum_{i=1}^{n_{s}}\sum_{j=1}^{n_{s}}\left(\mathcal{J}\left(C_{i},T\right)_{ij}-\hat{\mathcal{J}}\left(C_{i},T\right)_{ij}\right)^{2}},\label{eq:L2_fd}
\end{equation}
for four refinement levels using consecutively halved values of $\delta C$
and $\delta T$. The initial $\delta C$ is chosen as a quarter of
the smallest non-zero concentration, and the initial temperature perturbation
is chosen as $\delta T=100$ K. Figure~(\ref{fig:OOA}) shows the
order of convergence for a single random chemical state determined
using the UCSD chemical model~\citep{SanDiegoMech}. The refinement
results agree well with the expected second-order accuracy. We repeat
this refinement study across all previously mentioned chemical models
and observed consistent order of accuracy for both Jacobians (not
shown for brevity). In some cases, the initial $\delta C$ is too
small, leading to errors approaching machine precision and preventing
further reduction in the $L_{2}$-norm. By adjusting the initial $\delta C$,
the expected order of accuracy was recovered.

Additionally, we evaluate the computational cost of the analytical
Jacobian relative to their corresponding finite difference approximations.
Using the same approach to generate random chemical states as described
in Equation~(\ref{eq:chemical_state_01}), we compute 1,000 Jacobians and compared the total execution time. For the discrete
Jacobians, we use first-order finite difference schemes to minimize
computational cost,

\begin{eqnarray}
\mathcal{\hat{J}}_{ij,1} & = & \begin{cases}
\frac{\omega_{T}(C,T+\delta T)-\omega_{T}(C,T)}{\delta T} & i=1;\quad j=1\\
\frac{\omega_{T}(C_{j-1}^{+},T)-\omega_{T}(C,T)}{\delta C} & i=1;\quad j>1\\
\frac{\omega_{i-1}(C,T+\delta T)-\omega_{i-1}(C,T)}{\delta T} & i>1;\quad j=1\\
\frac{\omega_{i-1}\left(C_{j-1}^{+},T\right)-\omega_{i-1}\left(C,T\right)}{\delta C} & i>1;j>1.
\end{cases}\label{eq:finite_difference_Tevolve-1}
\end{eqnarray}
The results are summarized in Table \ref{tab:test} for a variety
of chemical models ranging in size and complexity. Despite using the
fastest discrete Jacobian optimization, the analytical Jacobians are
consistently faster across all cases. As the chemical reaction model increases
in complexity, the reduction in cost of computing the analytical Jacobian
relative to the finite different approximation becomes greater.

\begin{table}[htbp]
\centering
\caption{Comparison of Jacobian Timings. All Jacobians are normalized by their respective finite difference Jacobian \label{tab:test}}
\begin{tabular}{lcccccc}
\toprule
 & FFCM-2 (H2) & Ó Connaire & Burke & GRI-Mech 3.0 & UCSD & FFCM-2 (Full) \\
\midrule
$\mathcal{J}_{ij}/\hat{\mathcal{J}_{ij,1}}$                     & 2.84 & 3.23 & 3.03 & 9.52  & 9.61  & 19.28 \\
\bottomrule
\end{tabular}
\end{table}

Finally, Niemeyer et al.~\citep{Nie17} reported error metrics that
can be used to verify Jacobian accuracy. These metrics are utilized
in \ref{subsec:jacobianErel} to provide additional confidence
in the correctness of ChemGen's Jacobian computations.

\subsection{Implicit time integration}

\begin{figure}[H]
\begin{centering}
\includegraphics[scale=0.75]{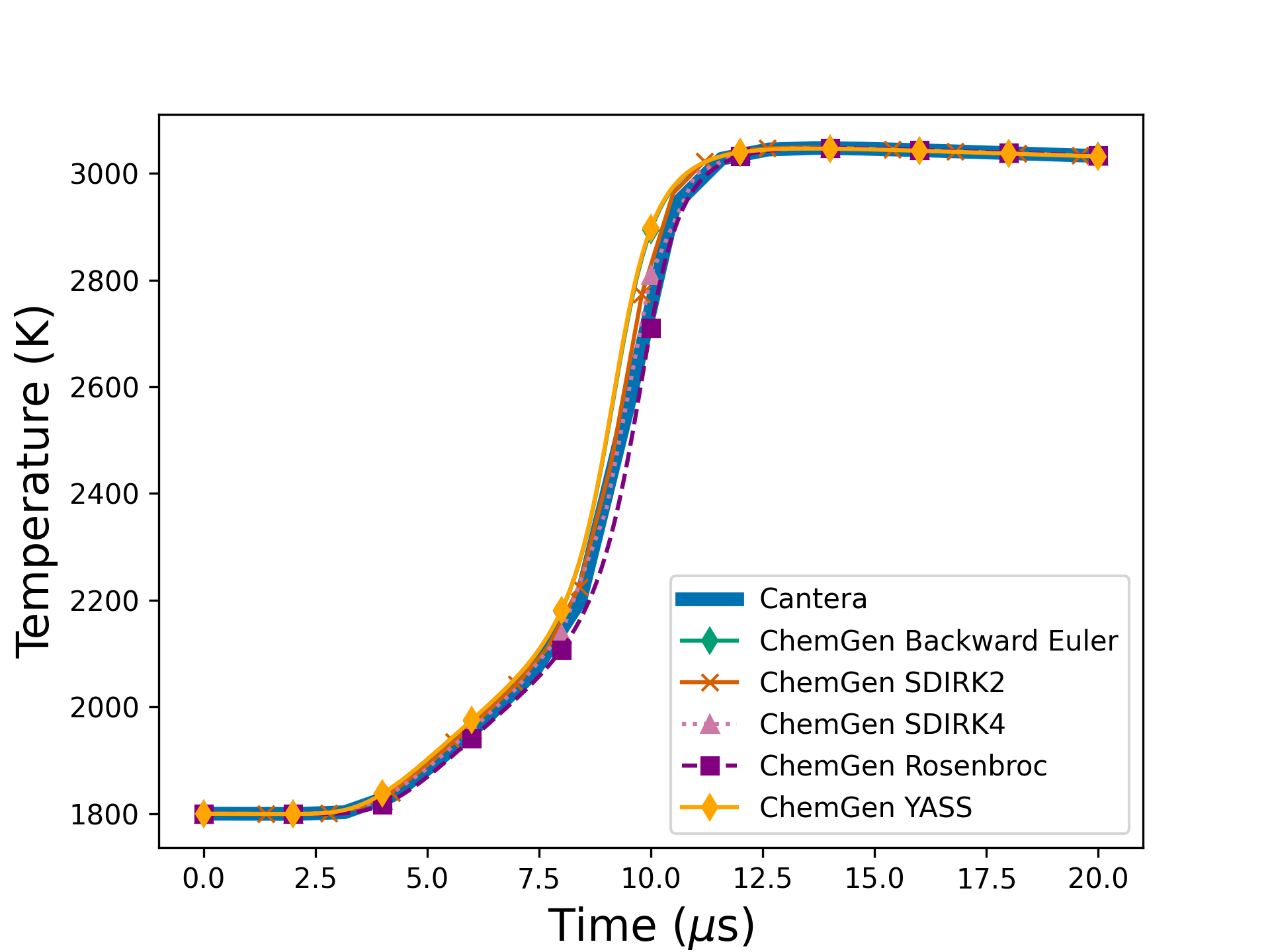}
\par\end{centering}
\caption{\label{fig:implicit_homogeneous_reactor}Homogeneous reactor results
obtained with ChemGen's implicit time integrators}
\end{figure}

This section presents results obtained with the implicit time integrators
provided by ChemGen, motivated by the confirmed accuracy of the source
term from Section~\ref{subsec:Source-Term-Accuracy}, successful
explicit time integration from Section~\ref{subsec:Explicit-Time-Integration},
and confidence in the computed source term Jacobian from Section~\ref{subsec:Jacobian-Verification}.
We utilize the FFCM-2 chemical model~\citep{Zha23} to integrate
a chemical state with initial conditions

\begin{eqnarray}
T & = & 1800\text{ K}\nonumber \\
p & = & 101325\text{ Pa}\nonumber \\
\left(X_{O_{2}},X_{N_{2}},X_{H_{2}},X_{C_{2}H_{4}}\right) & = & \left(0.2,0.4,0.2,0.2\right)\label{eq:chemical_state_03}
\end{eqnarray}
Figure~\ref{fig:implicit_homogeneous_reactor} compares the performance
of ChemGen's implicit time integrators against Cantera's ReactorNet
class. Among the integrators, the first-order methods, YASS and Backward
Euler, exhibit the largest discrepancies with Cantera's results. The
Rosenbrock method shows improved agreement, while SDIRK-2 and SDIRK-4
integrators match Cantera most closely.

The time step sizes are chosen (via trial and error) to yield the
fastest time-to-solution while maintaining reasonable accuracy. These
are summarized in Table ~(\ref{tab:time_step_and_timings}) with the Cantera simulation time, $\tau_{ct}$ normalized by the ChemGen simulation time, $\tau_{cg}$. The Cantera ReactorNet
time step is iterated on to find the fastest time to solution, while
utilizing both GMRES and the default adaptive preconditioner. ChemGen’s integration is often faster than Cantera’s, depending on the method. However, ChemGen is designed as a lean chemistry integrator, whereas Cantera may update additional state variables during the solve. Thus, Cantera’s integration should not be viewed as a direct chemistry-in-time tool, and some of the performance difference may reflect broader state management rather than solver efficiency alone.

\begin{table}[H]
\begin{centering}
\begin{tabular}{|c|c|c|}
\hline 
Integrator & $\Delta t$ & $\tau_{ct}/\tau_{cg}$\tabularnewline
\hline 
\hline 
Backward Euler & $10^{-7}$ s & $1.26$\tabularnewline
\hline 
SDIRK-2 & $7\times10^{-7}$ s & $3.19$\tabularnewline
\hline 
SDIRK-4 & $10^{-6}$ s & $3.51$\tabularnewline
\hline 
Rosenbroc & $10^{-7}$ s & $2.26$\tabularnewline
\hline 
YASS & $10^{-7}$ s & $3.22$\tabularnewline
\hline 
Cantera & $10^{-6}$ s & n/a\tabularnewline
\hline 
\end{tabular}
\par\end{centering}
\caption{Implicit time integrator optimal time step and simulation time\label{tab:time_step_and_timings}}
\end{table}

To assure the correct implementation of these implicit methods beyond
comparisons to Cantera, we assess order of convergence of the YASS,
SDIRK-4, and Rosenbrock methods. The implementation of SDIRK-2 is
similar to SDIRK-4 and Backward Euler is YASS with more linear solver
iterations. Therefore, we omit them from the discussion here. We employ
the extracted hydrogen submodel from FFCM-2 in a homogeneous reactor
simulation with initial conditions

\begin{eqnarray}
T & = & 1800\text{ K}\nonumber \\
p & = & 101325Pa\nonumber \\
\left(X_{O_{2}},X_{H_{2}},X_{N_{2}}\right) & = & \left(0.2,0.2,0.6\right).\label{eq:chemical_state_04}
\end{eqnarray}
We utilize the state at time $t=4\;\mu$s, which is the approximate
location of maximum $\frac{\partial T}{\partial t}$. This location
in time represents the approximate location in the solution where
the chemical state changes the most. We selecte $\Delta t=2\times10^{-6}$
for SDIRK-4 and $\Delta t=1\times10^{-7}$ for YASS and Rosenbroc.
We perform five refinements that change the integration time step,
$h$, by a factor of $2$, $h=\Delta t/2^{r}$ for $r=0\dots4$. We
use the finest solution, corresponding to $h=\Delta t/16$, as the
reference solution when computing the error,

\begin{equation}
\epsilon_{h}=\sum_{i=1}^{n_{s}}\left(C_{h,i}-C_{\frac{\Delta t}{16},i}\right)^{2}.\label{eq:L2_species}
\end{equation}

\begin{figure}[H]
\begin{centering}
\includegraphics[scale=0.75]{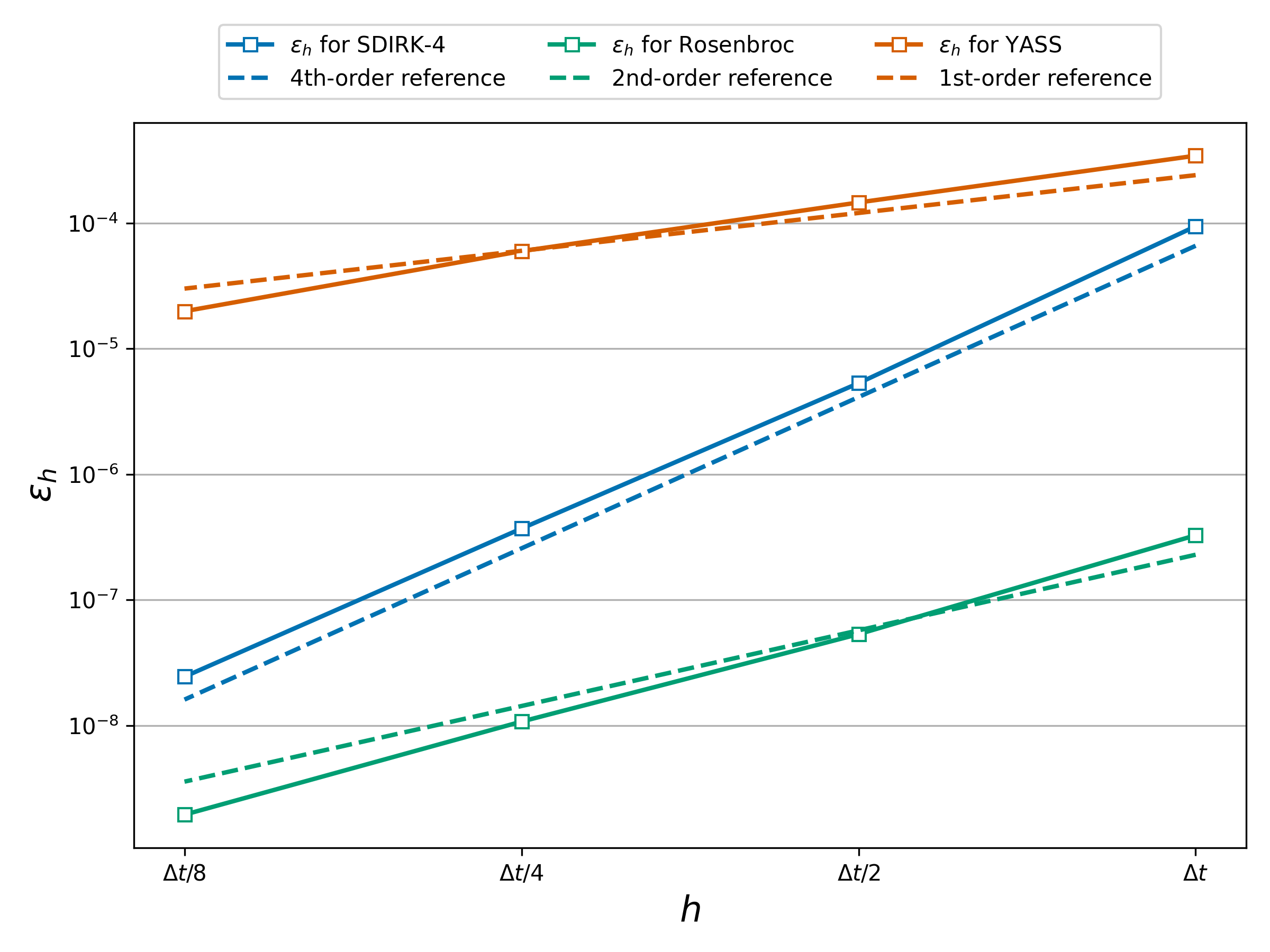}
\par\end{centering}
\caption{\label{fig:convergence_study}Convergence study for YASS, Rosenbroc,
and SDIRK-4 using a hydrogen submodel extracted from FFCM-2 with initial
conditions from Equation~\ref{eq:chemical_state_04}. The error is
evaluated at $t=4\,\mu\text{s}$. }
\end{figure}

\noindent The results of the convergence study are summarized in Figure~\ref{fig:convergence_study}.
The dashed lines represent the theoretical convergence rates. In all
three cases, the expected convergence rates are recovered.

\subsection{ChemGen in a detonation simulation}

To demonstrate the usage of ChemGen in an existing computational framework,
we target the chemistry evaluation in detonationFoam~\citep{Sun23}
which is based on the OpenFOAM package~\citep{Ope25}. In particular,
we focus on the detonationEulerFoam solver, which does not include
viscous effects. The solver treats each individual species conservation
equation via a segregated approach by solving

\begin{equation}
\frac{\partial\rho Y_{i}}{\partial t}+\nabla_{\mathrel{FVM}}\cdot\mathcal{F}\left(\rho Y_{i}\right)=\mathcal{R}_{i},\label{eq:openfom_fvm}
\end{equation}
where $\nabla_{\mathrel{FVM}}\cdot\mathcal{F}\left(\rho Y_{i}\right)$
denotes the finite volume discretization of the species flux and $\mathcal{R}_{i}$
is an applied source term of the $i^{th}$ specie. For the complete set
of conservation equations and additional details on the computational
strategy, see~\citep{Sun23}. Here the source term is calculated
as

\begin{equation}
\mathcal{R}=\left(\frac{C_{i}^{n+1}-C_{i}^{n}}{\Delta t}\right)W_{i}\label{eq:openfom_RR}
\end{equation}
and is held constant during each species equation solve. In Equation~\ref{eq:openfom_RR},
$W_{i}$ is the molecular weight, $C_{i}^{n}$ are the current concentrations
calculated from the current thermodynamic state, $\Delta t$ is the
desired fluid time step for integrating Equation~\ref{eq:openfom_fvm},
and $C_{i}^{n+1}$ are the predicted concentrations from solving the
species evolution ODEs, $\frac{\partial y_{c}}{\partial t}=S(y_{c})$.
Therefore, $\mathcal{R}_{i}$ is a linearized production term estimated
from solving the detailed kinetics.

In detonationFoam~\citep{Sun23}, the seulex ODE solver~\citep{Hai96}
supplied by OpenFOAM is used to solve for $C_{i}^{n+1}$. The seulex
solver utilizes an extrapolation algorithm based on the implicit Euler
method with step size control and order selection. Due to the small
time step size $\Delta t=1\times10^{-9}$ s we replace the seulex
solver with ChemGen's YASS integrator combined with Jacobi-preconditioned
GMRES.

We consider a two-dimensional detonation test that has been simulated
in several previous studies (see Oran et al.~\citep{Ora98}, Houim and
Kuo~\citep{Hou11}, Lv and Ihme~\citep{Lv15}, Johnson and Kercher~\citep{Joh20},
and Ching et al.~\citep{Chi22_2}). The domain here, which differs
slightly from those in the aforementioned works, is $0.5$~m long
($x_{1}$-direction) with a channel height of $0.06$~m ($x_{2}$-direction).
The initial mesh is a uniform quadrilateral mesh with $2000$ cells
in the $x_{1}$-direction and $240$ cells in $x_{2}$-direction.
We use the AMR default settings from~\citep{Sun23}, resulting in
three possible refinements, with a minimum mesh size of $31\times10^{-6}$
m. We use the slip wall condition for all boundaries. We again utilize
the hydrogen submodel extracted from FFCM-2~\citep{Zha23}. The initial
conditions are based on a steady detonation solution from the Shock
\& Detonation Toolbox~\citep{sdtoolbox}, where the pre-shock state
is given by

\begin{equation}
\begin{array}{ccc}
\qquad\qquad\qquad\qquad v_{1} & = & 0\text{ m/s},\\
\quad\quad\quad\;X_{Ar}:X_{O_{2}}:X_{H_{2}} & = & 7:1:2\\
\qquad\qquad\qquad\qquad p & = & 6670\text{ Pa}\\
\qquad\qquad\qquad\qquad T & = & 298\text{ K.}
\end{array}\label{eq:detonation-1d-initialization}
\end{equation}
In addition, we perturb the detonation front such that it varies in
the $x_{2}$-direction. In Sun et al.~\citep{Sun23} and in this
work a sinusoidal profile is specified to perturb the front location,
giving rise to instabilities that then transition to the expected
solution. 

\begin{figure}[H]
\begin{centering}
\includegraphics[width=6in]{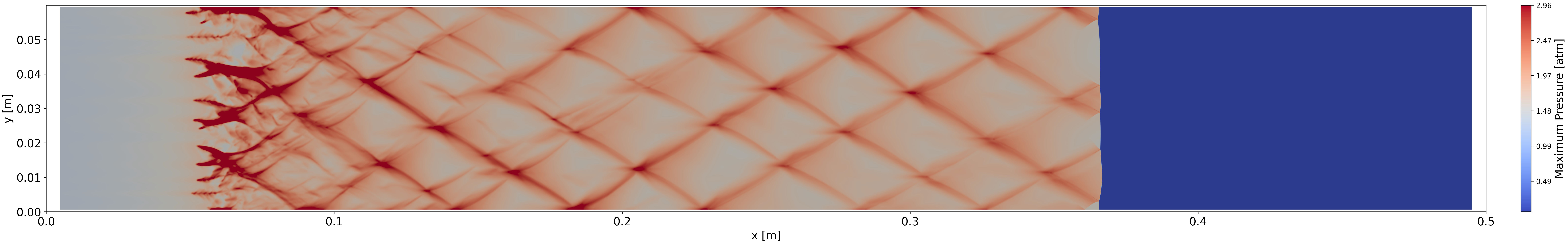}
\par\end{centering}
\caption{\label{fig:pressure-max-value}Maximum pressure, $p^{*,n+1}(x)=\max\left\{ p^{n+1}(x),p^{*,n}(x)\right\} $,
at $t=200\text{ }\mu$s.}
\end{figure}

The initial front progresses into the unreacted flow region, and the
irregular structures from the initialization cause the development
of two-dimensional fluid dynamic features eventually leading to an
established two-dimensional cellular structure. These structures were
visualized using a maximum-pressure history, $p^{*,n+1}(x)=\max\left\{ p^{n+1}(x),p^{*,n}(x)\right\} $
shown in Figure~\ref{fig:pressure-max-value} at $t=200\text{ }\mu$s.
These results were calculated by detonationEulerFoam using ChemGen
for the chemistry and demonstrate that ChemGen yields the expected
cellular detonation structures.

\begin{figure}[H]
\begin{centering}
\subfloat[Temperature distribution of detonation front at $t=200\text{ }\mu$s.
\label{fig:detonation_T}]{\begin{centering}
\includegraphics[width=0.45\linewidth]{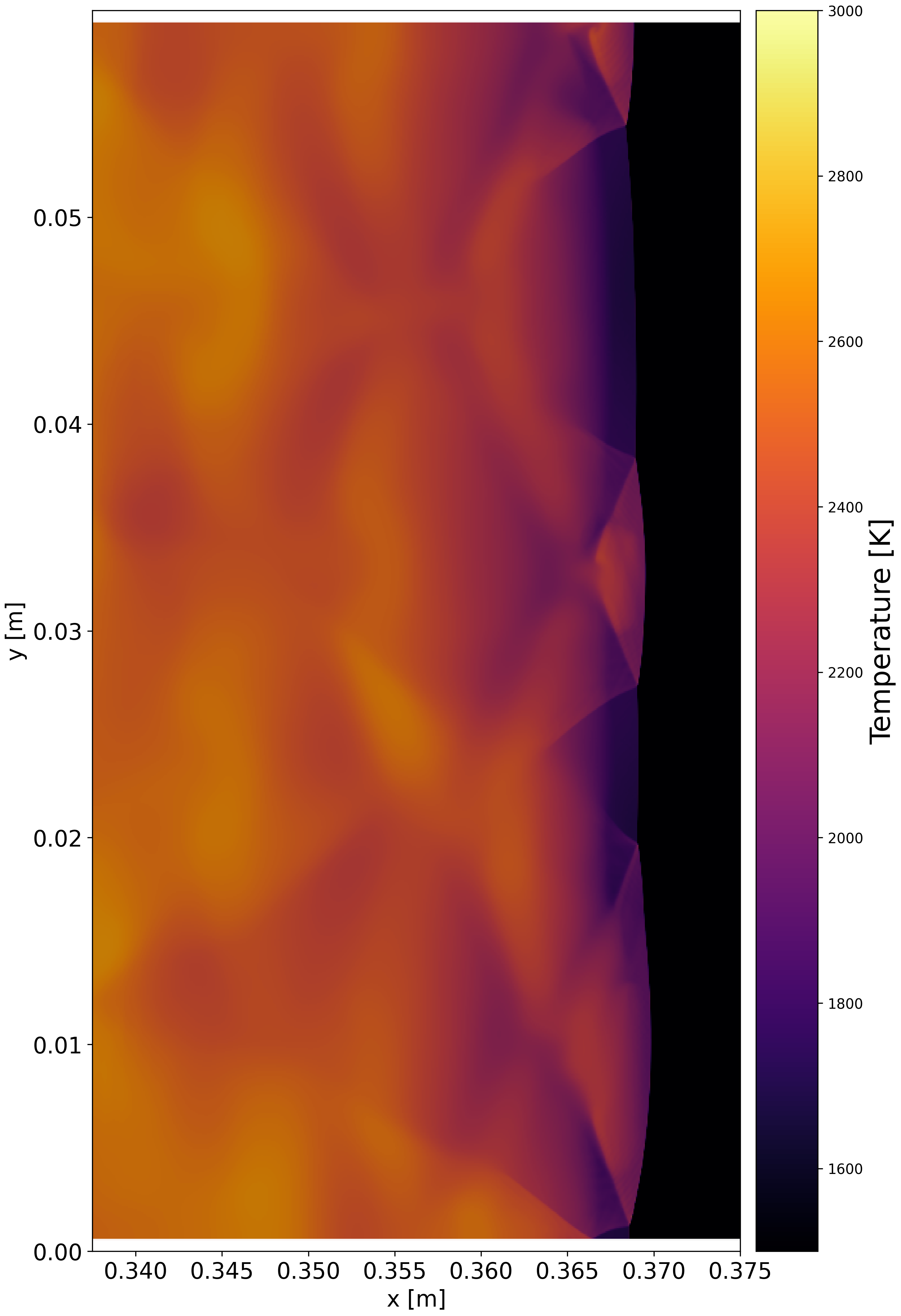}
\par\end{centering}
}\hfill{}\subfloat[Hydroxyl radical mass fraction, $Y_{OH}$, distribution of detonation
front at $t=200\text{ }\mu$s. \label{fig:detonation_OH}]{\begin{centering}
\includegraphics[width=0.45\linewidth]{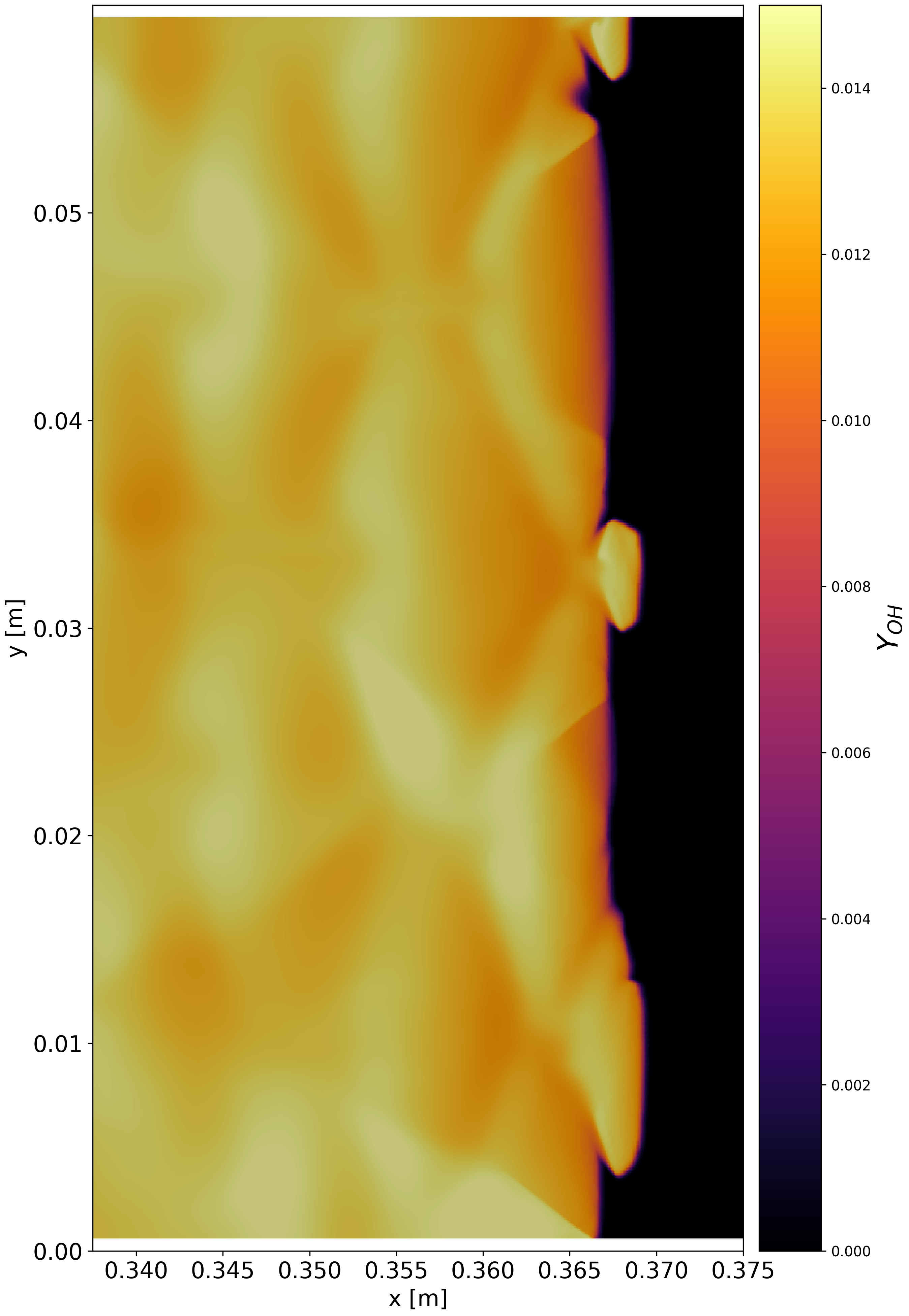}
\par\end{centering}
}
\par\end{centering}
\caption{\label{fig:detonation_T_OH} Temperature (left) and hydroxyl radical
mass fraction (right) distribution for two dimensional $Ar/H_{2}/O_{2}$
detonation using ChemGen to fully replace the chemistry solve in detonationEulerFoam.}
\end{figure}

\begin{figure}[H]
\begin{centering}
\subfloat[detonationEulerFoam reaction rate, $R_{of}$, for $O$-atom production
at $t=200\text{ }\mu$s.]{\begin{centering}
\includegraphics[width=0.45\linewidth]{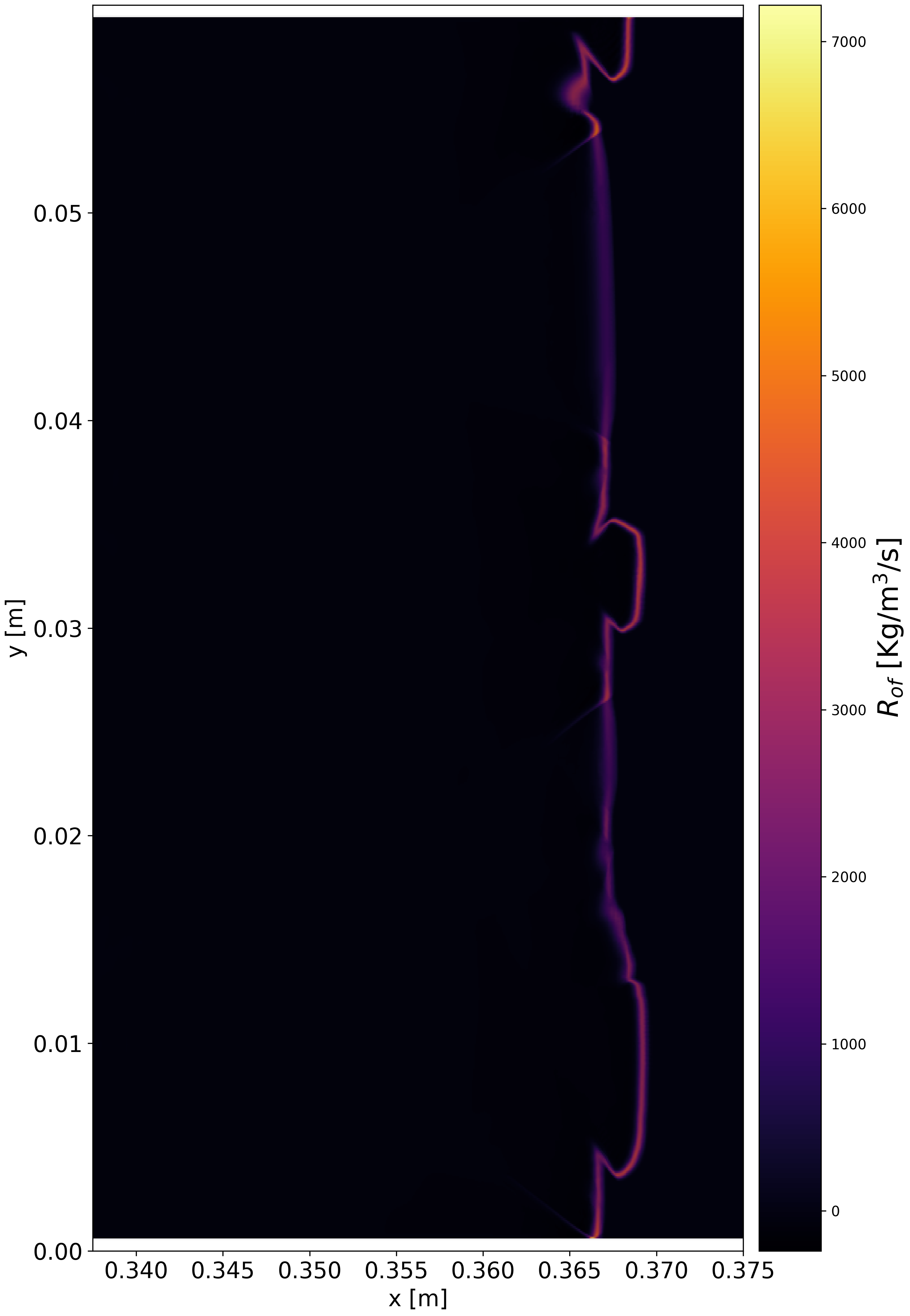}
\par\end{centering}
}\hfill{}\subfloat[\label{fig:RROFO}ChemGen reaction rate, $R_{cg}$, for $O$-atom
production at $t=200\text{ }\mu$s. \label{fig:RRCGO}]{\begin{centering}
\includegraphics[width=0.45\linewidth]{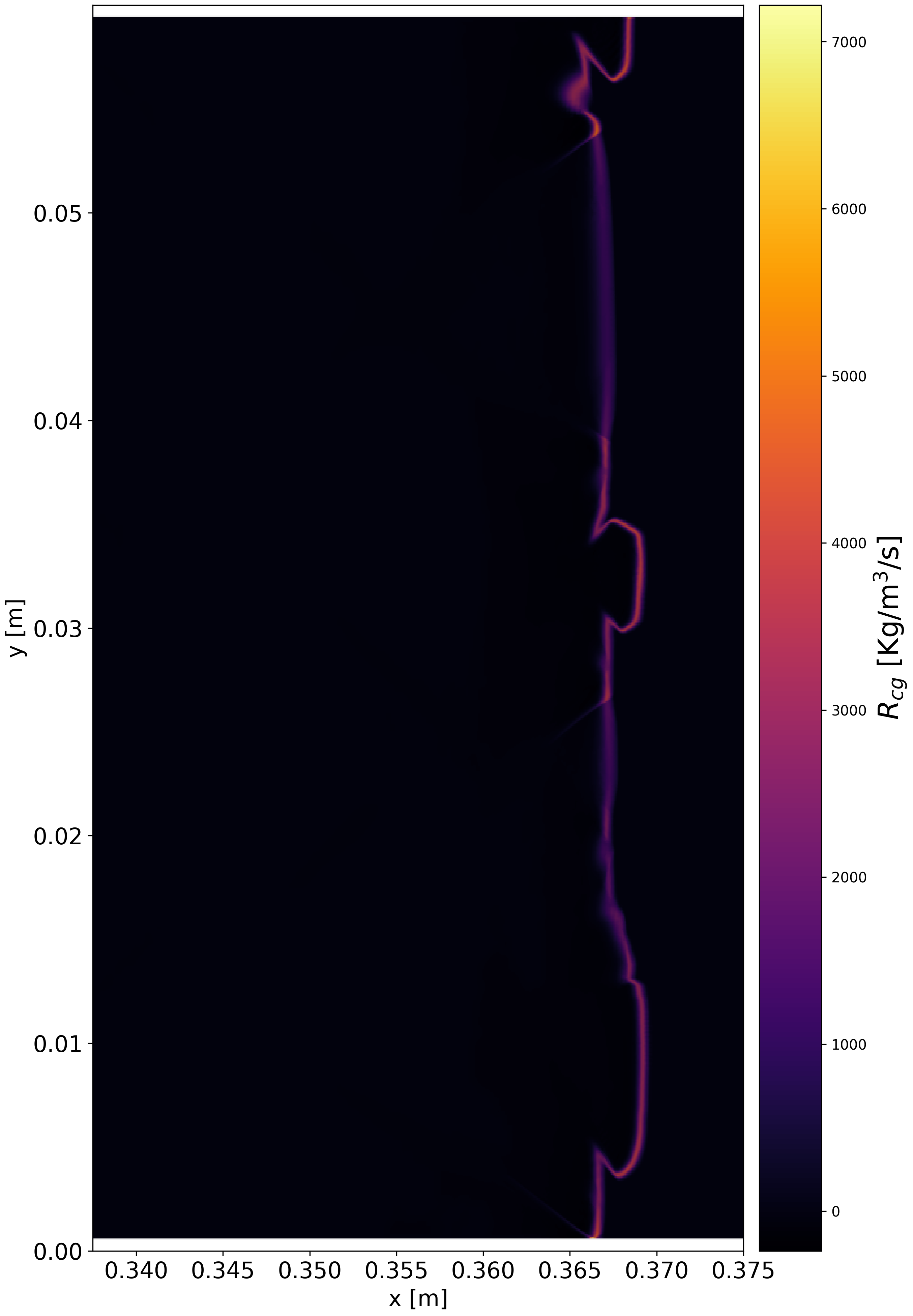}
\par\end{centering}
}
\par\end{centering}
\caption{\label{fig:reaction_rate_compare_OF_CG}The production term for $O$-atoms,
$\mathcal{R}=W_{i} \left(C_{O}^{n+1}-C_{O}^{n}\right)/ \Delta t $,
at $t=200\text{ }\mu$s for a detonationEulerFoam simulation of a
two dimensional detonation wave using OpenFOAM seulex ODE solver (left
with subscript $of$) and for ChemGen using the YASS ODE solver (right
with subscript $cg$).}
\end{figure}

Figures~\ref{fig:detonation_T} and~\ref{fig:detonation_OH} show
the distribution of temperature and hydroxyl radical mass fraction,
$Y_{OH}$, at $t=200\text{ }\mu$s. These results are consistent with
those produced using the existing seulex ODE integrator, which are
not reported here for brevity. Figures~\ref{fig:RRCGO} and~\ref{fig:RROFO}
show the production term for O-atoms, $\mathcal{R}_{O}=\left(\frac{C_{O}^{n+1}-C_{O}^{n}}{\Delta t}\right)W_{O}$,
at $t=200\text{ }\mu$s for detonationEulerFoam. The OpenFOAM result
of $\mathcal{R}_{O}$ using the seulex ODE solver is denoted as $R_{of}$
and the result using the ChemGen YASS ODE solver is denoted as $R_{cg}$
. These results demonstrate that an accurate CFD solution can be obtained
using ChemGen in the detonationEulerFoam solver. Furthermore, the
YASS solver provided by ChemGen is found to be on average 4 times
faster than the seulex solver.

\section{Conclusion}

In this work, we introduced ChemGen, a software package that leverages
code generation to integrate multispecies thermodynamics and chemical
kinetics into C\nolinebreak[4]\hspace{-.05em}\raisebox{.4ex}{\tiny\bf ++}-based computational physics software, with an
emphasis on CFD applications. ChemGen employs a decorator-based approach
to enable flexible and modular \Cpp code generation tailored to existing
simulation ecosystems. It generates both thermodynamic property approximations
and source term expressions, along with their analytical derivatives,
facilitating Jacobian assembly for implicit time integration. Additionally,
ChemGen includes a variety of time integrators, linear solvers, and
preconditioners.

We validated the correctness of the generated code by comparing its
results against established Cantera source terms and showed consistency
across a range of reaction models commonly used in CFD. We further
verified the accuracy of the Jacobian computations and demonstrated
optimal convergence across several supplied implicit integration methods.
Finally, we embedded ChemGen into an established CFD code to perform
a detonation simulation and reproduced results consistent with those
in the literature, illustrating its ability to integrate seamlessly
into mature computational frameworks. All data, source code, and supplementary
tutorials are publicly available.

ChemGen lays the groundwork for the consistent inclusion of chemical
source terms in diverse computational environments. Its flexibility
also opens the door for scalable solvers, reduced-order modeling strategies,
and novel chemistry integration methods that can advance
the frontier of high-fidelity chemical reacting flow simulation. ChemGen is an ongoing project released under the NRL Open
License, a source-available license provided by the U.S. Naval Research
Laboratory.

\section{Acknowledgment}
The work at Stanford is sponsored by Office of Naval Research under grants N00014-21-1-2475, N00014-22-1-2606 and N00014-23-1-2501. ESG and JEL also acknowledge support from the National Science Foundation under Grant No. DGE-1656518 and Grant No. DGE-2146755, respectively. RFJ, EJC, and JA acknowledge Dr. Eric Marineau of the Hypersonic Aerothermodynamics, High-Speed Propulsion and Materials Program of the Office of Naval Research Code 35 for support of this work.

\bibliographystyle{elsarticle-num}
\bibliography{citations,citations_runover}

\appendix

\section{Implicit time integrators\label{sec:Algorithms-of-Implicit}}

\subsection{SDIRK-2\label{subsec:SDIRK-2}}

The two stage Singly Diagonally Implicit Runge-{}-Kutta (SDIRK-2)
method belongs to a class of implicit Runge-Kutta schemes~\citep{Hai96,Fran97}.
These methods are stable and exhibit higher order of accuracy than
backward Euler, but are more expensive. The SDIRK-2 method is as follows:

\begin{eqnarray}
\Gamma_{1} & = & S\left(y_{n}+\gamma\Delta t\Gamma_{1}\right)\label{eq:SDIRK2}\\
\Gamma_{2} & = & S\left(y_{n}+(1-\gamma)\Delta t\Gamma_{1}+\gamma\Delta t\Gamma_{2}\right)\nonumber \\
y_{n+1} & = & y_{n}+\Delta t\left[(1-\gamma)\Gamma_{1}+\gamma\Gamma_{2}\right]\nonumber 
\end{eqnarray}
For the first stage,

\begin{equation}
f\left(\Gamma_{1}\right)=\Gamma_{1}-S\left(y_{n}+\gamma\Delta t\Gamma_{1}\right)
\end{equation}

\begin{equation}
\delta\Gamma_{1}=\left(\Gamma_{1,k+1}-\Gamma_{1,k}\right)=-\mathcal{G}\left(\Gamma_{1,k}\right)^{-1}f(\Gamma_{1,k})\label{eq:newtons-method-sdirk2}
\end{equation}

\begin{equation}
\mathcal{G}\left(\Gamma_{1,k}\right)\delta\Gamma_{1}=-f(\Gamma_{1,k})\label{eq:newtons-method-sdirk2-1}
\end{equation}
where

\begin{equation}
\mathcal{G}_{ij}\left(\Gamma_{1,k}\right)=\delta_{ij}-\gamma\Delta t\mathcal{J}_{ij}\left(y_{n}+\gamma\Delta t\Gamma_{1,k}\right)\label{eq:newtons-method-sdirk2-1-1}
\end{equation}
Here $\omega_{i}$ is evaluated from the initial guess of $\Gamma_{1,k=1}=\mathcal{S}\left(y_{n}\right)$
with subsequent $\Gamma_{1,k}$ coming from the iterated for decrement,
$\Gamma_{1,k+1}=\Gamma_{1,k}+\delta\Gamma_{1}$. Once $\Gamma_{1,k}$
is realized by the decrement being less than a prescribed tolerance,$||\delta\Gamma_{1,k}||<\epsilon$,
the next stage can be solved for in a similar manner,

\begin{equation}
f\left(\Gamma_{2}\right)=\Gamma_{2}-S\left(y_{n}+(1-\gamma)\Delta t\Gamma_{1}+\gamma\Delta t\Gamma_{2}\right)
\end{equation}

\begin{equation}
\delta\Gamma_{2}=\left(\Gamma_{2,k+1}-\Gamma_{2,k}\right)=-\mathcal{G}\left(\Gamma_{2,k}\right)^{-1}f(\Gamma_{2,k})\label{eq:newtons-method-sdirk2-2}
\end{equation}

\begin{equation}
\mathcal{G}\left(\Gamma_{2,k}\right)\delta\Gamma_{2}=-f(\Gamma_{2,k})\label{eq:newtons-method-sdirk2-1-2}
\end{equation}
Where

\begin{equation}
\mathcal{G}_{ij}\left(\Gamma_{2,k}\right)=\delta_{ij}-\gamma\Delta t\mathcal{J}_{ij}\left(y_{n}+(1-\gamma)\Delta t\Gamma_{1}+\gamma\Delta t\Gamma_{2,k}\right)\label{eq:newtons-method-sdirk2-1-1-1}
\end{equation}
Here $\omega_{i}$ is evaluated from the initial guess of $\Gamma_{2,k=1}=\mathcal{S}\left(y_{n}\right)$
with subsequent $\Gamma_{2,k}$ coming from the iterated for decrement,
$\Gamma_{2,k+1}=\Gamma_{2,k}+\delta\Gamma_{2}$. Once $\Gamma_{2,k}$
is realized the next stage can be solved for in a similar manner.
Chemgen utilizes $\gamma=1-\sqrt{2}/2$~\citep{Zha14}.

\subsection{SDIRK-4\label{subsec:SDIRK-4}}

The five stage Singly Diagonally Implicit Runge-Kutta (SDIRK-4) follows
the same structure as SDIRK-2 with additional stages requiring a set
of coefficients. We utilize the coefficients implemented by FATODE~\citep{Zha14}
which have shown stability for stiff problems and achieve 4th order
accuracy. The SDIRK-4 method is written as

\begin{eqnarray*}
y_{n+1} & = & y_{n}+\Delta t\sum_{i=1}^{5}b_{i}\Gamma_{i}\\
\Gamma_{p} & = & S\left(y_{n}+\sum_{j=1}^{i}a_{pj}\Delta t\Gamma_{j}\right)
\end{eqnarray*}
with coefficients

\begin{equation}
\begin{array}{ccc}
a_{ii}=0.2\bar{6} & a_{21}=0.5 & a_{31}=3.541539\times10^{-1}\\
a_{32}=-5.415395\times10^{-2} & a_{41}=8.515494\times10^{-2} & a_{42}=-6.484332\times10^{-2}\\
a_{43}=7.915325\times10^{-2} & a_{51}=2.100115 & a_{53}=2.399816\\
a_{54}=-2.998818 & b_{1}=2.100115 & b_{2}=-7.677800\times10^{-1}\\
b_{3}=2.399816 & b_{4}=-2.998818 & b_{5}=0.2\bar{6}
\end{array}
\end{equation}
The implicit strategy is similar to SDIRK-2 but written more generally
here

\begin{equation}
f\left(\Gamma_{p}\right)=\Gamma_{p}-S\left(y_{n}+\sum_{j=1}^{p}a_{pj}\Delta t\Gamma_{j}\right)
\end{equation}

\begin{equation}
\delta\Gamma_{p}=\left(\Gamma_{p,k+1}-\Gamma_{p,k}\right)=-\mathcal{G}\left(\Gamma_{p,k}\right)^{-1}f(\Gamma_{p,k})\label{eq:newtons-method-sdirk2-2-1}
\end{equation}

\begin{equation}
\mathcal{G}\left(\Gamma_{p,k}\right)\delta\Gamma_{p}=-f(\Gamma_{p,k})\label{eq:newtons-method-sdirk2-1-2-1}
\end{equation}
Where

\begin{equation}
\mathcal{G}_{ij}\left(\Gamma_{p,k}\right)=\delta_{ij}-a_{pp}\Delta t\mathcal{J}_{ij}\left(y_{n}+\sum_{j=1}^{p}a_{pj}\Delta t\left(\Gamma_{j}\right)_{k}\right).\label{eq:newtons-method-sdirk2-1-1-1-1}
\end{equation}
Like SDIRK-2, the subsequent $\Gamma_{p,k}$ comes from the iterated
for decrement, $\Gamma_{p,k+1}=\Gamma_{p,k}+\delta\Gamma_{p}$. Once
$\Gamma_{p,k}$ is realized by the decrement being less than a prescribed
tolerance,$||\delta\Gamma_{p}||<\epsilon$, the next stage can be
solved for in a similar manner.

\subsection{2nd-order Rosenbrock\label{subsec:Rosenbroc}}

Rosenbrock schemes are another attractive time integration strategy
which remove the need of multiple Jacobian inversions that are required
in a Newton solve per stage as shown in the prior SDIRK based methods.

\begin{eqnarray*}
\left(\frac{1}{\Delta t\gamma_{11}}\mathcal{I}-\mathcal{J}_{n}\right)\Gamma_{1} & = & S\left(y_{n}\right)\\
\left(\frac{1}{\Delta t\gamma_{22}}\mathcal{I}-\mathcal{J}_{n}\right)\Gamma_{2} & = & S\left(y_{n}+\alpha\Gamma_{1}\right)+\frac{\beta}{\Delta t}\Gamma_{1}-S\left(y_{n}\right)\\
y_{n+1} & = & y_{n}+m_{1}\Gamma_{1}+m_{2}\Gamma_{2}
\end{eqnarray*}
Here $\mathcal{J}_{n}$ is the chemical Jacobian, $\mathcal{J}_{ij}$,held
constant for both stages, thus removing the need for multiple Jacobian
evaluations. Again we utilize the coefficients from FATODE~\citep{Zha14},

\begin{equation}
\begin{array}{ccccc}
\gamma_{ii}=\gamma= & 1+\frac{\sqrt{2}}{2}\text{, }\alpha=1/\gamma\text{, } & \beta=\frac{-2}{\gamma} & m_{1}=\frac{3}{2\gamma} & m_{2}=\frac{1}{2\gamma}\end{array}.
\end{equation}

\subsection{YASS\label{subsec:YASS}}

Yet another stiff solver (YASS) is an appealing solver that only uses
one linear solve to predict the species evolution. YASS is equivalent
to one Newton step for backwards Euler that relies on sub-stepping
if the change in $y_{n}$ is too large. The integration is

\begin{equation}
\left(\mathcal{I}-\Delta t\mathcal{J}\right)\delta y_{n}=\Delta tS\left(y_{n}\right)
\end{equation}
where $y_{n+1}=y_{n}+\delta y_{n}$. If the norm of the decrement
is larger than some scalar value, $||\delta y_{n}||>\alpha$, then
the YASS integration's time step is reduced to $\Delta t_{s+1}=\frac{1}{2}\Delta t_{s}$
where $\Delta t_{s}=\Delta t$ in the first integration. This means
that there is an appealing chance that the solver is performing a
minimum of one linear solve. This necessitates restricting the solver
to fail if the time step reaches a minimum value, $\Delta t_{s+1}<\Delta t_{\text{min}}$
, to where a possible higher order method can be utilized. 

\section{Constant pressure strategy\label{subsec:Constant-Pressure-Strategy}}

In compressible CFD, it is typically assumed that the volume of a
cell remains constant over time. This assumption influences the allowable
time integration strategies and, in compressible flows, permits pressure
to rise in response to changes in species. However, in incompressible
or low-Mach solvers, pressure is treated differently: it is assumed
to adjust instantaneously, effectively decoupling it from the time
evolution and thus remain constant at some thermodynamic value. This
means pressure is held constant in time, which imposes a constraint
on the energy conservation. Specifically, if pressure does not evolve,
then the enthalpy, $\rho h=\sum_{i=1}^{n_{s}}W_{i}C_{i}h_{i}$ , must
also remain constant in time. This constraint becomes clear when examining
the thermodynamic relationship between internal energy, enthalpy,
and pressure,

\begin{equation}
\rho u=\sum_{i=1}^{n_{s}}W_{i}C_{i}u_{i}=\sum_{i=1}^{n_{s}}W_{i}C_{i}h_{i}-R^{o}T\sum_{i=1}^{n_{s}}C_{i}=\sum_{i=1}^{n_{s}}W_{i}C_{i}h_{i}-p=\rho h-p,\label{eq:int_energy_enthalpy}
\end{equation}
and applying a time derivative

\begin{equation}
\frac{\partial\rho u}{\partial t}=\frac{\partial\rho h}{\partial t}-\cancelto{0}{\frac{\partial p}{\partial t}}=\frac{\partial\rho h}{\partial t}=0.\label{eq:int_energy_enthalpy-1-1}
\end{equation}

Therefore, the constant pressure assumption can be enforced by treating
enthalpy as a globally constant, while leaving the rest of the system
formulation unchanged. In the same manner that arrived at Equation~(\ref{eq:temperature_source})
a source term for temperature is found under constant enthalpy,

\begin{equation}
\frac{\partial\rho h}{\partial t}=\frac{\partial\rho h}{\partial T}\frac{\partial T}{\partial t}+\sum_{i=1}^{n_{s}}\frac{\partial\rho h}{\partial C_{i}}\frac{\partial C_{i}}{\partial t}=0\label{eq:temperature_source_h-1}
\end{equation}
which gives

\begin{equation}
\frac{\partial T}{\partial t}=\omega_{T}=-\frac{\sum_{i=1}^{n_{s}}W_{i}h_{i}\omega_{i}}{c_{p}}.\label{eq:temperature_source_h}
\end{equation}

\section{Source term and source term Jacobian errors}

\subsection{Source term error due to thermodynamic polynomial degree\label{subsec:sourcetermpoly}}

\begin{figure}[H]
\begin{centering}
\includegraphics[scale=0.4]{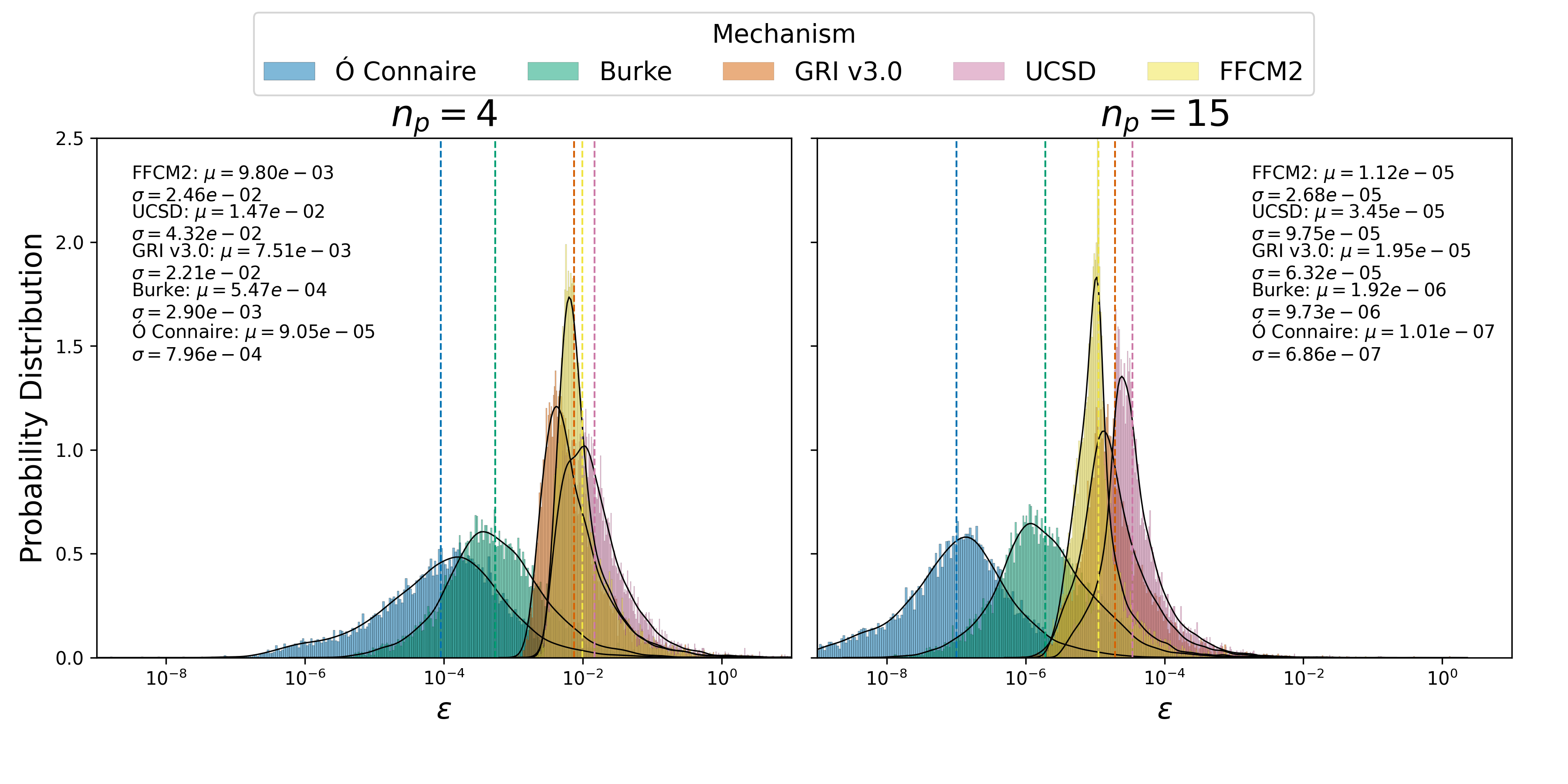}
\par\end{centering}
\caption{\label{fig:source_term_accuracy_np}Error distributions (Equation~(\ref{eq:source_error}))
for ChemGen source calculations in comparison to Cantera source calculations
for five tested chemical models: Ó Conaire \citep{Oco04}, Burke \citep{Bur12},
GRI-Mech 3.0 \citep{Smi00}, UCSD \citep{SanDiegoMech}, and FFCM2
\citep{Zha23}. The vertical dash line represents the mean of the
distributions and is colored according to the corresponding model.
The distribution mean, $\mu$, and corresponding standard deviation,
$\sigma$, are also reported.}
\end{figure}

To investigate the effect of polynomial order on source term accuracy,
we perform the same error analysis of Section~(\ref{subsec:Source-Term-Accuracy})
while varying $n_{p}$. Figure~(\ref{fig:source_term_accuracy_np})
presents the results for two alternative choices: $n_{p}=4$ (left)
and $n_{p}=15$ (right). The error distributions shift with polynomial
order. Specifically, the mean errors $\mu$ associated with $n_{p}=4$
are approximately an order of magnitude larger than those for $n_{p}=7$,
while the $n_{p}=15$ results exhibit mean errors that are an order
of magnitude smaller. The standard deviation $\sigma$ follows the
same trend, indicating that increasing the polynomial degree not only
improves the accuracy but also enhances the precision of the predicted
source terms.

The reduction in error with increasing polynomial degree provides
additional confidence in the correctness of the ChemGen implementation.
We also found that several parameters in the UCSD model, particularly
those used in Troe fall-off reactions, contain extremely large or
small values. For instance, in reaction 9, the parameters are specified
as \texttt{T3: 1.0e-30, T1: 1.0e+30} . The interpretation and handling
of such extreme values can vary between software implementations.
Examining $F_{\mathrm{cent}}$, given by

\[
F_{\mathrm{cent}}(T)=(1-\alpha_{j})\exp(-T/T_{j,3})+\alpha\exp(-T/T_{j,1})+\exp(-T_{j,2}/T),
\]
we note that the very small value of $T_{3}$ effectively drives the
first term to zero, while the extremely large value of $\ensuremath{T_{1}}$
drives the second exponential term to one. It is assumed, based on
the specification, that omitting $T_{2}$ entirely implies that the
third term is excluded from the calculation, rather than setting $\ensuremath{T_{2}=0}$~\citep{cantera}.
Updating ChemGen to address these edge cases and avoid numerical instabilities
led to reduced error and recovered the expected trend that higher-order
polynomial fits improved accuracy.

\subsection{Source term Jacobian error \label{subsec:jacobianErel}}

Niemeyer et al. verified the PyJac-generated Jacobian by comparing
their analytical Jacobian to a differentiated Jacobian across a wide
range of thermochemical states, evaluating between 100,000 and 1 million
samples depending on the chemical model~\citep{Nie17}. They reported
two metrics that could be used per chemical state to evaluate the
relative error,

\begin{equation}
E_{rel}=||\frac{\mathcal{J}_{ij}-\hat{\mathcal{J}_{ij}}}{\hat{\mathcal{J}_{ij}}}||{}_{F}=\sqrt{\sum_{i=1}^{n_{s}}\sum_{j=1}^{n_{s}}\left(\frac{\mathcal{J}\left(C_{i},T\right)_{ij}-\hat{\mathcal{J}}\left(C_{i},T\right)_{ij}}{\hat{\mathcal{J}_{}}\left(C_{i},T\right)_{ij}}\right)^{2}}\text{and}\label{eq:E_rel}
\end{equation}

\begin{equation}
E_{norm}=\frac{\|\mathcal{J}_{ij}-\hat{\mathcal{J}_{ij}}\|{}_{F}}{\hat{\|\mathcal{J}_{ij}}\|{}_{F}}=\frac{\sqrt{\sum_{i=1}^{n_{s}}\sum_{j=1}^{n_{s}}\left(\mathcal{J}_{}\left(C_{i},T\right)_{ij}-\hat{\mathcal{J}}\left(C_{i},T\right)_{ij}\right)^{2}}}{\sqrt{\sum_{i=1}^{n_{s}}\sum_{j=1}^{n_{s}}\left(\hat{\mathcal{J}}\left(C_{i},T\right)_{ij}\right)^{2}}}.\label{eq:E_norm}
\end{equation}
In these error metrics, only matrix elements where $|\mathcal{J}_{ij}|>\hat{||\mathcal{J}_{b}}||_{F}/10^{20}$
are considered, where the subscript $F$ denotes the Frobenius norm,
i.e., $||A||_{F}=\sum_{j=1}^{m}\sum_{i=1}^{n}\left(A_{ij}\right)^{2}$
for matrix$A$ of size $n\times m$.

\begin{figure}[H]
\begin{centering}
\includegraphics[scale=0.4]{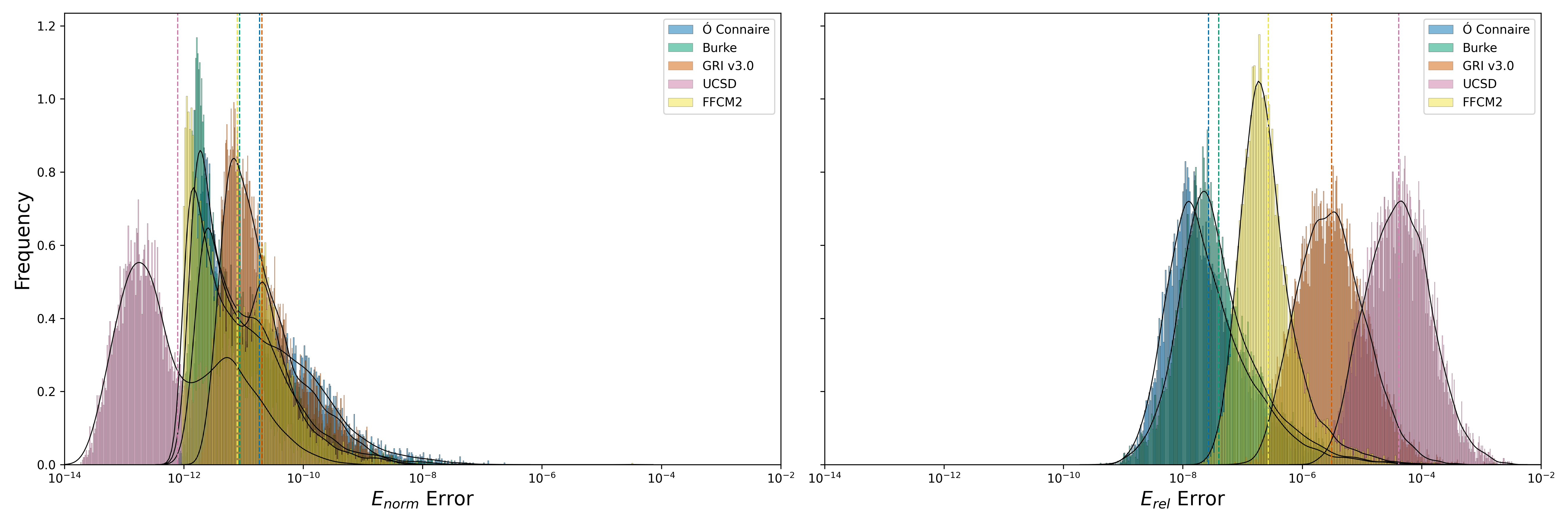}
\par\end{centering}
\caption{\label{fig:jacobian_error_distribution} Distributions of $E_{nom}$
and $E_{rel}$ for ChemGen source term Jacobian calculations in comparison
to Cantera source calculations for five tested chemical models from
Table~\ref{tab:chemical models}}
\end{figure}

In Niemeyer et al., it was reported that $E_{rel}$ may be more useful
in identifying discrepancies in both larger and smaller matrix elements.
Table~(\ref{tab:jacobian_norms}) reports the mean and maximum error
norms for both $E_{rel}$ and $E_{norm}$ using ChemGen for five chemical
models with 10,000 randomly generated chemical states. Each analytical
Jacobian, $\mathcal{J}_{ij}$, is compared to the second order finite
difference Jacobian, $\hat{\mathcal{J}_{ij}}$, with $\delta C=10^{-6}$
Kmol/m$^{3}$and $\delta T=10^{-3}$ K. Overall, both error norms
suggest that the analytical Jacobian is producing the expected derivatives.
In this study, $E_{rel}$ errors are generally higher, but never exceeding
$3\times10^{-3}$. Figure~(\ref{fig:jacobian_error_distribution})
shows the distributions of $E_{norm}$ (left) and $E_{rel}$ (right).
The $E_{norm}$ distributions are orders of magnitude smaller than
the $E_{rel}$ distributions. Regardless, these results give confidence
in the ability to accurately calculate the analytical Jacobians.

\begin{table}
\caption{$E_{norm}$ and $E_{rel}$ mean and max errors for different chemical
models given 10,000 randomly produced states each. Overall, the $E_{rel}$
errors were generally higher, but never exceeding $3\times10^{-3}$.\label{tab:jacobian_norms}}

\begin{tabular}{|c|c|c|c|c|}
\hline 
Model & $E_{norm}$ Mean Error & $E_{norm}$ Max Error & $E_{rel}$ Mean Error & $E_{rel}$ Max Error\tabularnewline
\hline 
\hline 
Ó Connaire & $3.92\times10^{-10}$ & $2.39\times10^{-7}$ & $3.38\times10^{-7}$ & $8.71\times10^{-5}$\tabularnewline
\hline 
Burke & $1.23\times10^{-10}$ & $5.43\times10^{-8}$ & $4.44\times10^{-7}$ & $1.01\times10^{-4}$\tabularnewline
\hline 
GRI v3.0 & $8.46\times10^{-11}$ & $5.57\times10^{-9}$ & $8.10\times10^{-6}$ & $4.07\times10^{-4}$\tabularnewline
\hline 
UCSD & $7.55\times10^{-12}$ & $7.27\times10^{-10}$ & $9.06\times10^{-5}$ & $2.53\times10^{-3}$\tabularnewline
\hline 
FFCM-2 & $3.41\times10^{-9}$ & $3.35\times10^{-5}$ & $2.70\times10^{-4}$ & $1.70\times10^{-4}$\tabularnewline
\hline 
\end{tabular}
\end{table}

\end{document}